\documentclass{aa}

\usepackage{lscape}
\usepackage{rotating}
\usepackage{wrapfig}
\usepackage{graphicx}
\usepackage{txfonts}

\usepackage[normalem]{ulem}
\begin{document}

   \title{VLT near- to mid-IR imaging and spectroscopy of the M\,17 UC1\,--\,IRS5 region \thanks{Based on observations by the European Southern Observatory Very Large Telescope on Cerro Paranal, Chile (ESO program IDs: 281.C-5027(A), 281.C-5051(A,B))}}

   \author{Zhiwei Chen\inst{1,2,3} \and Dieter E.~A. N{\"u}rnberger~\inst{2} \and  Rolf Chini\inst{2,4} \and Zhibo Jiang\inst{1} \and Min Fang\inst{1}}

   \institute{Purple Mountain Observatory \& Key Laboratory for Radio Astronomy, Chinese Academy of Sciences, 2 West Beijing Road, 210008 Nanjing, China \\
\email{zwchen@pmo.ac.cn}
\and Astronomisches Institut, Ruhr--Universit{\"a}t Bochum, Universit{\"a}tsstrasse 150, 44801 Bochum, Germany \and University of Chinese Academy of Sciences, 100039 Beijing, China \and Instituto de Astronom{\'i}a, Universidad Cat{\'o}lica del Norte, Avenida Angamos 0610, Casilla 1280 Antofagasta, Chile
             }

   \date{Received date; accepted date}

  \abstract
{
}
   {We investigate the surroundings of the hypercompact \ion{H}{ii} region M\,17 UC1 to probe the physical properties of the associated young stellar objects and the environment of massive star formation.}
   {We use diffraction-limited near-IR (VLT/NACO) and mid-IR (VLT/VISIR) images to reveal the different morphology at various wavelengths. Likewise we investigate the stellar and nebular content of the region by VLT/SINFONI integral field spectroscopy with a resolution $R\sim1500$ at $H+K$ bands. }
   {Five of the seven point sources in this region show $L$-band excess emission. Geometric match is found between the H$_2$ emission and near-IR polarized light in the vicinity of \object{IRS5A}, and between the diffuse mid-IR emission and near-IR polarization north of \object{UC1}. The H$_2$ emission is typical for dense PDRs, which are FUV pumped initially and repopulated by collisional de-excitation. The co-presence of \ion{He}{i},  \ion{H}{i}, and H$_2$ lines in most region argues against an edge-on configuration of the M\,17 SW PDR, but is in favor of a moderately inclined geometry with respect to the line of sight. The spectral types of \object{IRS5A} and \object{B273A} are B3$-$B7 V/III and G4$-$G5 III, respectively. The observed infrared luminosity $L_\mathrm{IR}$ in the range $1-20\,\mu$m is derived for three objects; we obtain $2.0\times10^3\,L_\sun$ for IRS5A, $13\,L_\sun$ for IRS5C, and $10\,L_\sun$ for B273A.}
   {\object{IRS5} might be a young quadruple system. Its primary star \object{IRS5A} is confirmed to be a high-mass protostellar object ($\sim 9\,M_\sun, \sim 1\times10^5\,\mathrm{yrs}$); it might have terminated accretion due to the feedback from the stellar activities (radiation pressure, outflow) and the expanding \ion{H}{ii} region of M\,17. \object{UC1} might also have terminated accretion because of the expanding hypercompact \ion{H}{ii} region ionized by itself. The disk clearing process of the low-mass YSOs in this region might be accelerated by the expanding \ion{H}{ii} region. The outflows driven by UC1 are running in south-north with its northeastern side suppressed by the expanding ionization front of M\,17; the blue-shifted outflow lobe of IRS5A is seen in two types of tracers along the same line of sight in the form of H$_2$ emission filament and mid-emission. The H$_2$ line ratios probe the properties of M\,17 SW PDR, which is confirmed to have a clumpy structure with two temperature distributions: warm, dense molecular clumps with $n_\mathrm{H}>10^5\,\mathrm{cm}^{-3}$ and $T\approx575\,$K and cooler atomic gas with $n_\mathrm{H}\sim3.7\times10^3-1.5\times10^4\,\mathrm{cm}^{-3}$ and $T\sim50-200$\,K.}


   \keywords{stars: early-type --stars: formation --stars: individual: M\,17 UC1 --ISM: individual objects: M\,17 SW --ISM: molecules: photon-dominated region (PDR)}

   \maketitle
%

\section{Introduction}
Massive stars~($> 8\,M_{\sun}$) affect their surroundings by ionizing radiation and strong stellar winds throughout their life, as well as by metal enrichment in their final fate as Supernovae. However, the formation process of massive stars is still a hot open debate. Both theoretical prediction and observational evidence suggest that massive stars most likely form via accretion of material quite similar to low-mass stars, but the involved processes may not be only simply scaled up~\citep[][and references therein]{2007ARA&A..45..481Z}. The major difference is the energetic feedback~(radiation pressure, stellar winds and outflows) which rapidly dissipates the circumstellar envelope and consequently limits the mass growth of the central young stellar object~(YSO). In addition, for high-mass YSOs the earliest evolutionary phase (protostellar phase, before reaching the zero-age main sequence, hereafter ZAMS) is very short, as characterized by Kelvin-Helmholtz timescales of less than $10^5$ yrs. Thus, massive stars evolve fast, even during their accretion phase. It is thought that the time spent in the main accretion phase might be significantly less than the main-sequence lifetime, presumably of the order of a few dynamical times of the star-forming molecular core. Therefore, to accumulate sufficiently large amounts of material during the very short timescale, accretion rates of high-mass YSOs must be much higher than those of low-mass YSOs, e.g., $\gtrsim10^{-4}\,M_{\sun}\,\mathrm{yr}^{-1}$~\citep[e.g.][hereafter H+10]{2010ApJ...721..478H} compared with $< 10^{-8} \,M_{\sun}\,\mathrm{yr}^{-1}$~\citep[e.g.][]{2013A&A...549A..15F}. The strong feedback from massive protostars may prevent the accretion even before the arrival at the ZAMS if accretion rates are constantly larger than a few $10^{-3}\,M_{\sun}\,\mathrm{yr}^{-1}$. This settles an upper limit for the protostars around several tens of solar masses. In fact, accretion rates may vary strongly with the evolution of massive protostars. Massive accretion might continue in a non-steady fashion, which potentially allows even more massive stars to form by mass accretion~(H+10).

The evolution of massive protostars at such high accretion rates is still the vague feature of massive star formation which determines the feedback on their environment. Numerical simulations show that accreting massive protostars with high accretion rates have large radius. For instance, at an accretion rate of $1\times10^{-3}\,M_{\sun}\,\mathrm{yr}^{-1}$, the protostellar radius may exceed $100\,R_{\sun}$ at maximum \citep[e.g.][]{2008ASPC..387..189Y,2009ApJ...691..823H}. Such a large radius hence leads to low effective temperature, and very low stellar UV luminosity, which might be too low for the growth of an \ion{H}{ii} region around the protostar~\citep{2002ARA&A..40...27C,2007ApJ...666..976K,2010MNRAS.405.1560M}. High-mass protostars are considered to precede the formation of an \ion{H}{ii} region. From the observational view, a lot of high-mass protostar candidates have high infrared luminosities without observable \ion{H}{ii} regions~\citep{2007A&A...472..155K,2008A&A...481..345M,2009A&A...498..147G}. 

Sect.~\ref{Sec:uc1} briefly introduces the massive star-forming region M\,17, and particularly summarizes the studies of the M\,17 UC\,--\,IRS5 region. In Sect.~\ref{Sec:OBS}, the observations and data reduction are described. The results based on near- to mid-IR data are presented in Sect.~\ref{Sec:Re}. The derived properties are discussed in Sect.~\ref{Sec:Di}, and the conclusions are presented in Sect.~\ref{Sec:Co}.

\section{M\,17 UC1\,--\,and the southwestern photodissociation region \label{Sec:uc1}}

In a distance of $1.98_{-0.12}^{+0.14}$\,kpc~\citep{2011ApJ...733...25X}, \object{M\,17} is among the best laboratories in the Galaxy for investigating the formation of massive stars. In this paper, we report near- to mid-IR imaging and integral-field spectroscopic studies for the M\,17 UC1\,--\,IRS5 region~(see Fig.~\ref{Fig:Fc}), which is located just west of the arc-like ionization front (IF). This region is well known because of the hypercompact \ion{H}{ii}~(HCHIIregion \object{M\,17~UC1}~\citep{2004ApJ...605..285S}, which is surrounded by a circumstellar disk~\citep[][hereafter N+07]{2007ApJ...656L..81N}. The other interesting object, \object{M\,17 IRS5}, is a bright IR source located 5$\arcsec$ southwest of \object{M\,17~UC1}; in contrast to the HCHII region it is not detectable at 1.3\,cm~\citep{2000A&A...357L..33C}. Its spectral energy distribution~(SED) at IR wavelengths suggests a warmer component with color temperature $\sim1000$\,K, and a cooler component $\sim150$\,K~\citep[][hereafter N+01]{2001A&A...377..273N}. The non-detection of an associated \ion{H}{ii} region is reminiscent of the early protostellar phase when the protostar is huge with low effective temperature and low UV luminosity -- meaning that \object{IRS\,5} might be younger than UC1. However, this explanation is just one of the three plausible scenarios proposed by \citet{2002AJ....124.1636K} (hereafter K+02). An \ion{H}{ii} region with a density higher than $3\times10^5$\,cm$^{-3}$ or a heavy dusty envelope could also escape detection. Previous near-IR polarization studies revealed infrared reflection nebulae (IRN) associated with the two sources~\citep[][hereafter CZ+12]{2012PASJ...64...110C}, which might trace potential outflows.  The third bright IR source, \object{M\,17 B273}, projected against the edge of the arc-shaped IF, might be a YSO too \citep[near-IR excess, see][]{1997ApJ...489..698H}, but requires a spectroscopic classification. 

Besides characterizing the stellar content, the SINFONI integral-field spectroscopy can also be used to investigate the diffuse nebular emission of M\,17 SW\,--\,one of the best-studied dense PDRs in the Galaxy. At far-IR to millimeter wavelengths, studies of molecular and atomic emission indicate that the structure of the gas is highly clumped \citep{1988ApJ...332..379S,1990ApJ...356..513S,1992ApJ...390..499M,2010A&A...510A..87P}, and supported by magnetic field rather than by thermal gas pressure \citep{2007ApJ...658.1119P}. Temperatures of $\sim275\,\mathrm{K}$ were found toward the IF \citep{2001ApJ...560..821B}. One characteristic of PDRs is H$_2$ emission originating from the collisional deexcitation of H$_2$ molecules initially excited by UV photons, which is an important heating mechanism of dense PDRs \citep{1989ApJ...338..197S}.  \citet{2013ApJ...774L..14S} reported four mid-IR pure-rotational H$_2$ lines toward M\,17 SW which are consistent with H$_2$ emission in high-density clumps ($n_\mathrm{H}>10^5\,\mathrm{cm}^{-3}$) embedded in an interclump atomic gas of density lower by two or three orders of magnitude \citep{1992ApJ...390..499M, 1993ApJ...405..216M}.  \citet{2007A&A...465..931N} reported near-IR H$_2$ emission associated with a jet ejected by a forming high-mass protostar, which is just located $1\arcmin$ southeast to the M\,17 UC1\,--\,IRS5 region. In such case, H$_2$ emission is produced by thermal emission in shock fronts.


\section{Observations and data reduction\label{Sec:OBS}}


\subsection{SINFONI observations}
 \label{Sec:OBS-SINFO}

\begin{figure*}
\centering
\includegraphics[width=0.7\textwidth]{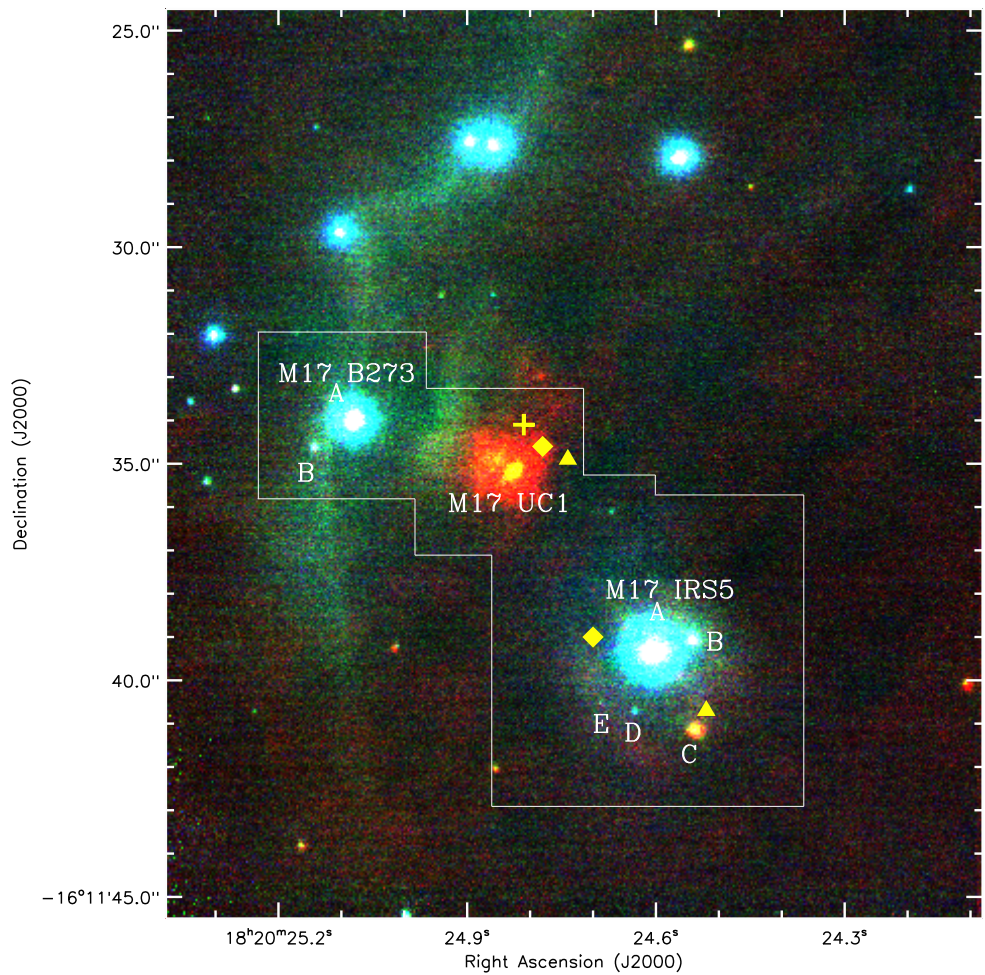}
\caption{Three-color image of the M\,17 SW PDR taken by NACO at three near-IR broadband filters (blue: $H$; green: $K$; red: $L$), with the area of the SINFONI integral field spectroscopy outlined in the white box. Individual point sources discussed in this paper are labeled. Three types of maser are marked according to their coordinates: 22-GHz water masers \citep[filled triangles,][]{1998MNRAS.297..215C,1998ApJ...500..302J}; Class II methanol masers at 6.66 GHz \citep[filled diamonds,][]{1995MNRAS.272...96C,2000MNRAS.313..599C}; OH masers at 1.67 GHz \citep[thick plus,][]{1998MNRAS.297..215C}.}
\label{Fig:Fc}

\end{figure*}

  High spatial resolution (AO supported), middle spectral resolution ($R\sim1500$) 
  near-IR integral field spectroscopic data of the \object{M\,17 UC1\,--\,IRS5} region (see Fig.~\ref{Fig:Fc}) 
  were taken in Service Mode during the nights 2008-06-08/09, 2008-09-26/27 and 
  2008-09-27/28, using ESO's near-IR integral field spectrograph SINFONI \citep[{\em {\bf S}pectrograph for {\bf IN}tegral {\bf F}ield {\bf O}bservations in the 
  {\bf N}ear-{\bf I}nfrared};][]{2005Msngr.120...26G} mounted on the Cassegrain focus 
  of the VLT Yepun at the Paranal observatory, Chile. 

\begin{table*}
\centering

\caption{SINFONI observation parameters}
\begin{tabular}{c c c c c}
\hline\hline
Target & R.A. & DEC. & Exp. Time & Telluric STD  \\
          & (J2000) & (J2000) & (s)        &               \\   
\hline
M17-IRS5 & 18 20 24.60 & $-$16 11 39.4 & $27\times60$  & HIP\,091126 (G2\,V) \\
         & 18 20 24.71 & $-$16 11 37.1 & $6\times60$   & HIP\,094378 (B5\,V) \\
M17-UC1  & 18 20 24.82 & $-$16 11 34.9 & $6\times60$   & HIP\,094378 (B5\,V) \\
M17-B273 & 18 20 25.07 & $-$16 11 33.9 & $18\times60$  & HIP\,092470 (B2\,IV) \\

\hline
\end{tabular}
\label{Tbl:Obs}
\end{table*}

 Overall, we observed a sequence of 4 adjacent target positions (see Table \ref{Tbl:Obs}). In all cases, AO curvature sensing was performed on the reference source M17-CEN\,64 (RA\,$=$\,18:20:25.71, DEC\,$=$\,$-$16:11:41.7, J2000), which is located at the distance of about $16\farcs5$ toward the east of IRS5 (about $12\arcsec$ toward the south-east of B273). For appropriate sky subtraction, a nearby empty sky position was available at the distance of about $70\arcsec-90\arcsec$ toward the west of IRS5. 

 In a trade-off between avoiding saturation of the spectra of the central point source and optimizing both dynamic range and field of view (FOV) for the diffuse circumstellar emission, we set up SINFONI with its 100\,mas pixel scale, covering an instantaneous FOV of about 3$\arcsec$\,$\times$\,3$\arcsec$ with 64\,$\times$\,32\,pixels of 50\,mas\,$\times$\,100\,mas each, together with the $H+K$ grating, which comprises the wavelength range from about 1.45\,$\mu$m to 2.45\,$\mu$m at the spectral resolution of about 1.0\,nm (dispersion of 0.50\,nm/pixel). 
  The detector integration times (DITs) were set to 60\,s (with NDIT\,$=$\,1), both for on-source (science) exposures and off-source (sky) exposures. 

  For each target position, we applied a sequence of dither offsets on a 3\,$\times$\,3 
  position grid with grid spacings of 0$\farcs$5, resulting in an effective FOV of 
  about $4\arcsec\times4\arcsec$. 
  Because the faint circumstellar material around IRS5 is rather widespread, we took 
  additional exposures on a 3\,$\times$\,3 position grid with grid spacings of 2$\farcs$0 
  and 2$\farcs$2, increasing the effective FOV to about $7\farcs4\times7\farcs4$. 
  In total, we gathered 27 exposures on IRS5, 6 exposures on the position intermediate 
  between IRS5 and UC1, 6 exposures on UC1 and 18 exposures on B273. 
  To allow proper sky subtraction, science exposures (with the AO loop closed, of course) 
  were interleaved every 10 minutes by several (typically 5-6) exposures on the sky position 
  (although void of stars and diffuse emission, with the AO loop open). 

  To correct for telluric features and to flux-calibrate the science exposures, each night 
  they were immediately followed by observations of a suitable telluric standard star, 
  applying the same strategy (auto-jitter pattern, fixed sky offset) as for the 
  science targets and matching their airmass. The telluric standard stars used in the three observation nights are listed in Table \ref{Tbl:Obs}.

  During all three nights the observing conditions were good, with sky transparency clear 
  (exception: thin cirrus clouds passing through during the night 2008-09-27/28) and 
  telescope guide probe seeing measurements typically in the range $0\farcs6-1\farcs2$. Correspondingly, the achieved strehl ratios were in the range $20-40\%$, as directly 
  measured on the spectrally collapsed data cubes. 

  Dark frames, lamp flats, and arcs (taken with SINFONI's internal Neon and Argon lamp for 
  the purpose of wavelength calibration) were obtained through the SINFONI scientific 
  calibration plan. 
  All basic steps of data reduction (flat-fielding, sky subtraction, bad pixel and 
  atmospheric distortion correction) as well as the wavelength calibration were performed 
  with the SINFONI data reduction pipeline. 
  Final merging of the fully reduced and wavelength calibrated individual data cubes to 
  one mosaic covering all science targets was performed within the SINFONI pipeline, too. 

\subsection{Ancillary near- to mid-IR data \label{Sec:OBS-IMG}}
The $JHKL$ AO imaging was carried out in 2003 June using NAOS/CONICA~\citep[NACO;][]{2003SPIE.4841..944L,2003SPIE.4839..140R} on the ESO VLT at the Paranal observatory, Chile. The FOV is $27\arcsec\times27\arcsec$ with pixel resolution of $0\farcs027$. The 3$\sigma$ limiting magnitudes are $J=20.2$, $H=19.7$, $K=19.3$, and $L=15.2$. The photometry was carried out merely for the point sources within the FOV of SINFONI data (see Fig.~\ref{Fig:Fc}). The magnitudes of these point sources were extracted based on a variety of aperture sizes on the purpose of aperture correction. The final magnitudes with aperture correction are calibrated with photometric standard stars, which are \object{HD\,110621} for $J$-band ($J=8.91$), \object{HD\,188112} for $HK$-band ($H=10.78$, $K=10.89$), and \object{HD\,161743} for $L$-band ($L=7.61$). The astrometry is adjusted by referencing the NACO point sources with the detections of the SINFONI data whose astrometry was calibrated using the 2MASS catalog. With this procedure, the relative astrometric difference between the NACO data and SINFONI data is better than $0\farcs1$.

The TIMMI2 \citep[{\em {\bf T}hermal {\bf I}nfrared {\bf M}ultimode {\bf I}nstrument};][]{1998SPIE.3354..865R} mid-IR imaging was carried out in 2003 July at the ESO 3.6\,m telescope at La Silla, Chile. The observations covered the $N1$, $N10.4$, and $Q1$ bands with $\lambda_\mathrm{eff}$ of $8.7\,\mu$m, $10.38\,\mu$m, and $17.72\,\mu$m, respectively, and all had FOV of $55\arcsec \times 38\arcsec$ with pixel scale of $0\farcs2$. All data are limited by diffraction with a FWHM (Full Width Half Maximum) of $0\farcs7$.

The M\,17 UC1\,--\,IRS5 region was imaged with VISIR \citep[{\em {\bf V}LT {\bf I}mager and {\bf S}pectrometer for mid-{\bf I}nf{\bf R}ared};][]{2004Msngr.117...12L} in 2006 May through the Sic filter ($\lambda_\mathrm{eff}=11.85\,\mu$m). The observation procedures are described in N+07. The image is of good quality (FWHM\,$\approx 0\farcs32$) limited by diffraction, with pixel scale of $0\farcs127$. The astrometry of the TIMMI2 and VISIR imaging data were calibrated on the basis of the NACO data. The astrometric accuracy is better than $0\farcs1$ throughout the FOV centered on the M\,17 UC1\,--\,IRS5 region.

The spectroscopy of the $N$-band silicate absorption feature was performed with TIMMI2 at the ESO 3.6\,m telescope at La Silla, Chile, within the same observation run of \object{UC1} (N+07). The seeing was $0\farcs7$; the slit width was $1\farcs2$.

\begin{table*}
\caption{NACO $JHKL$ photometry of the point sources in \object{M\,17 UC1\,--\,IRS5} region}
 \centering
  \begin{tabular}{c c c c c c c}    
\hline\hline
    Source & R.A.    & DEC   & $J$  & $H$ &$K$& $L$ \\
      ID     &(J2000)&(J2000)&(mag)&(mag)&(mag)&(mag)\\
   \hline 
                                                                                       M\,17 B273 \\       
    \object{A}& $18~20~25.08$ & $-16~11~34.0$ & $14.03\pm0.01$ &$12.19\pm0.01$ & $10.71\pm0.01$ & $8.89\pm0.01$  \\
    \object{B} & $18~20~25.14$ & $-16~11~34.6$ &  $20.11\pm0.33$  & $17.18\pm0.05$ & $15.26\pm0.04$ & $13.25\pm0.09$ \\
    \object{~M\,17 UC1} & $18~20~24.84$ &$-16~11~35.1$ & $-$  & $-$ &$13.61\pm0.02$ & $6.20\pm0.03$ \\
     M\,17 IRS5 \\
    \object{A} & $18~20~24.60$ & $-16~11~39.4$ &$13.34\pm0.01$&$11.38\pm0.02$&$9.81\pm0.01$ & $8.16\pm0.03$    \\
    \object{B} & $18~20~24.54$ & $-16~11~39.0$ &$19.07\pm0.22$&$15.09\pm0.02$&$13.81\pm0.02$& $11.56\pm0.03$   \\
    \object{C} & $18~20~24.54$ & $-16~11~41.2$ & $-$     &$19.13\pm0.13$&$14.65\pm0.03$& $9.69\pm0.01$   \\
    \object{D} & $18~20~24.63$ & $-16~11~40.7$ & $-$      &$17.97\pm0.12$&$16.32\pm0.07$& $15.27\pm0.40$   \\
    \object{E} & $18~20~24.69$ & $-16~11~40.5$ & $-$&$19.69\pm0.32$ &$17.75\pm0.18$& $-$   \\
    \hline
  \end{tabular}
\tablefoot{Coordinates of point sources adopted here are a compromise between the astrometry of NACO and SINFONI data, whose differential has been checked less than 0\farcs1 throughout the FOV.}
\label{Tbl:Ph}
\end{table*}   

\section{Results\label{Sec:Re}}

\subsection{High angular Resolution Near-IR imaging \label{Sec:NIRIMA}}
The \object{M\,17 UC1\,--\,IRS5} region outlines a typical interface between an \ion{H}{ii} region and a PDR, as characterized by a prominent IF seen in our previous, lower resolution near-IR images~\citep[e.g.][CZ+12]{2008ApJ...686..310H}. Our high-resolution images unveil many fine structures toward this region which were not revealed before~(see Fig.~\ref{Fig:Fc}). Part of the results had been published for UC1 (N+07). Interestingly, the two luminous infrared objects (\object{M\,17 IRS5}, \object{M\,17 B273}) are resolved to have more than one component in the high-resolution near-IR images. If they actually are multiple systems, one needs to review these objects because they were treated as single high-mass YSO candidates, and thus the primary's brightness would have been overestimated. 

The $JHKL$ magnitudes of the objects inside the white box are listed in Table~\ref{Tbl:Ph}. The brightest component of the \object{IRS5} system is treated as the primary star, namely \object{IRS5A}, and followed by four other fainter companions \object{IRS5B}, \object{IRS5C}, \object{IRS5D}, and \object{IRS5E}. Similarly, \object{B273A} is chosen as the primary star of the binary \object{B273}.

\subsubsection{Point sources with IR excess}
\label{Sec:IRE}

\begin{figure}
\centering

\resizebox{\hsize}{!}{\includegraphics{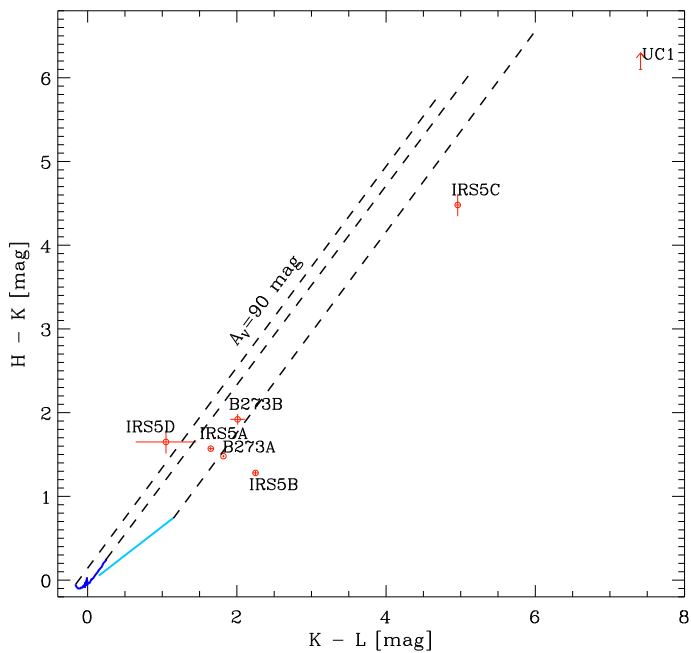}}
\caption{NACO $K-L$ versus $H-K$ color-color diagram. The blue curve denotes the intrinsic colors of main-sequence stars \citep{2001ApJ...558..309D}; the cyan line corresponds to the locus of T Tauri stars \citep{1997AJ....114..288M}. Reddening vectors (dashed lines) with slope of $E(H-K)/E(K-L)=1.2$ \citep{2008ApJ...686..310H} are drawn. Their lengths equal to visual extinction of 90 mag.}
\label{Fig:Ccd}
\end{figure}

Fig.~\ref{Fig:Ccd} shows the $HKL$ color-color diagram of the objects from Table~\ref{Tbl:Ph}. Objects located to the right of the reddening vectors are suggested to show infrared excess which traces the circumstellar material around YSOs. The level of infrared excess somehow relates to the amount of circumstellar material \citep{1992ApJ...393..278L}, which in turn reflects the evolutionary stage of YSO. All the objects with accurate $L$-band magnitudes show infrared excess, indicating that they are YSOs. \object{IRS5A}, \object{B273A} and \object{B273B} all show infrared excess resembling to classical T Tauri stars (CTTS). \object{IRS5B} seems to have a larger infrared excess, implying a younger evolutionary stage. \object{IRS5C} is the reddest source apart from \object{UC1}, although its infrared excess is comparable to that of \object{IRS5B}. Without proper spectral type classification, we can only establish a crude mass sequence for these point sources according to their dereddened magnitudes. However, it is impossible to extract exact extinction values for these sources solely based on their locations in Fig.~\ref{Fig:Ccd}. The close positions in the $HKL$ color-color diagram suggest that all sources except \object{IRS5C} have similar reddening. The extinction of \object{IRS5C}, however, is about three times higher. From the $H$-band magnitudes, we suggest that \object{IRS5A} is most massive source, followed by \object{B273A} and \object{IRS5C}, \object{IRS5E} has the lowest mass while the other sources have masses in between. 

The area enclosed by the white outline harbors several types of maser, such as water (H$_2$O) maser at 22 GHz, hydroxyl (OH) maser at 1.665 GHz and Class II methanol maser at 6.66 GHz. Such a variety of maser is commonly found toward regions of massive star formation \citep[e.g.][]{2010A&A...517A..56F,2010MNRAS.406.1487B}. Particularly, the Class II methanol maser is believed to be radiatively pumped, and typically coincides in position with hot molecular cores, UC\ion{H}{ii} regions, OH masers and near-IR sources \citep{2010A&A...517A..56F}. Note that H$_2$O, OH, and 6.66-GHz methanol masers are all detected in the close vicinity of \object{UC1}, which coincides with the expectation of a massive forming star. Moreover, a 6.66-GHz methanol maser is detected in close proximity to \object{IRS5A}. The H$_2$O maser at 22 GHz is closer to \object{IRS5C} than to \object{IRS5A}. If the 22-GHz H$_2$O maser is associated with \object{IRS5C}, it might be a high-mass YSO too. The other possibility is that the 22-GHz H$_2$O maser is associated with \object{IRS5A}, coexisting with the 6.66-GHz methanol maser. 


\subsubsection{Diffuse emission}
\label{Sec:En}
A bright rim crossing between the \object{B273} binary is clearly seen in Fig.~\ref{Fig:Fc}. This rim is a portion of the IF in \object{M\,17} seen in larger view (e.g. CZ+12) that represents the boundary between the \ion{H}{ii} region (to the northeast) and \object{M\,17 SW} (to the southwest). We note an elongated, bar-like emission feature $2\arcsec$ west to the IF. This emission bar is nearly parallel to the IF, and shows similar brightness and color with the IF. However, the emission bar has a smooth boundary while the IF is very sharp. Hence, the emission bar might not have the same physical origin as the IF.

\begin{figure}
\centering
\resizebox{\hsize}{!}{\includegraphics{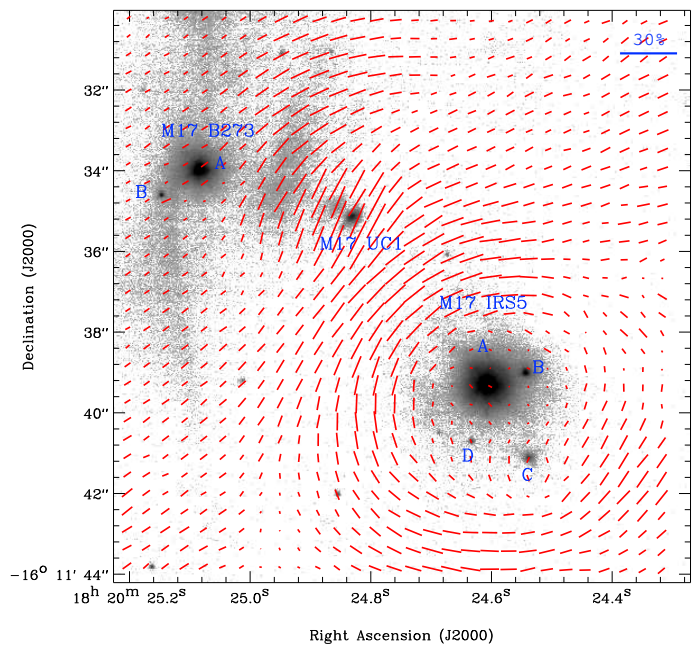}}
\caption{IRSF/$K$ polarization vectors (red lines; CZ+12) overlaid on NACO/$K$ image (grey scale). }
\label{Fig:Pol}
\end{figure}

The near-IR polarization studies toward M\,17 (CZ+12) revealed a bar-like feature enhanced in $K$-band polarized light lying west to the IF. Fig.~\ref{Fig:Pol} shows the comparison between $K$-band polarization and NACO/$K$ image. The polarization pattern centered on \object{IRS5A} had been discussed in CZ+12. The high polarization degree of \object{UC1} is consistent with its two reflection lobes (N+07). At the position of the emission bar seen in NACO/$K$ image, a concentration of polarization vectors $\sim10\%-15\%$ marks the location of an IRN. From the pattern of these polarization vectors, \object{UC1} is the best candidate for the illuminating source of the IRN.  In fact, the coincidence the two bar-like features seen in different manners indicates a mass concentration of gas and dust associated with \object{UC1} in a bar-like shape. 

\subsection{Mid-IR imaging }
\label{Sec:Mir}
\begin{figure}
\centering
\resizebox{\hsize}{!}{\includegraphics{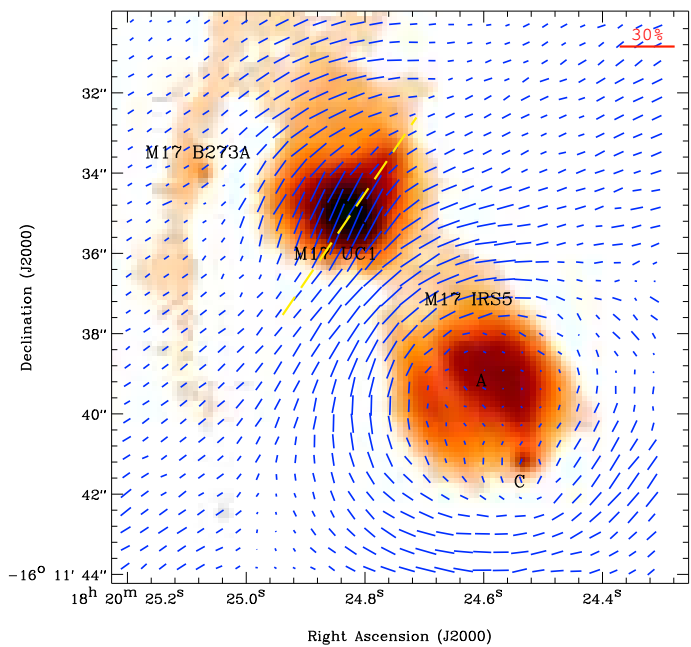}}
\caption{VISIR $11.85\,\mu$m image of the M\,17 UC1--IRS5 region. IRSF/$K$ polarization vectors (blue lines; CZ+12) are overplotted as well. The disk position angle is denoted by the yellow dashes with P.A of $156\degr$ (N+07).}
\label{Fig:Mir}
\end{figure}

Previous mid-IR imaging showed round morphologies both for \object{UC1} and \object{IRS5} with angular resolution $1-2\arcsec$. Interestingly, the new VISIR image at 11.85\,$\mu$m with angular resolution $\approx0\farcs3$ reveals substructures seen in Fig.~\ref{Fig:Mir} for both objects. The IF is still visible at this wavelength, but much fainter than in the $K$-band.

\object{UC1} shows an elongated feature toward northwest, which agrees well with its circumstellar disk orientated at P.A. (from north to east) $\approx146\degr$ (shown in yellow dashes in Fig.~\ref{Fig:Mir}). Moreover, the two parts separated by the circumstellar disk are asymmetric, with the northeastern part more extended. This asymmetric structure can be explained by an inclined circumstellar disk with angle $\approx30\degr$ with respect to the line of sight (hereafter LOS), which is proposed by N+07 to reproduce the $K$-band asymmetric structure which possesses a much brighter southwestern lobe. Besides the mid-IR emission in the close vicinity of \object{UC1}, a structure extending to north is more worthy to note. Interestingly, it is spatially coinciding with the IRN discussed in Sect.~\ref{Sec:En}, and fully covers the area of substantial $K$-band polarization. Conversely, the bar-like feature
seen in $K$-band covers only a portion of the IRN. 

The high-sensitivity VISIR image shows two lobes separated by a dark lane orientated at P.A. $\approx30\degr$ for \object{IRS5A}. Unlike \object{UC1}, the $K$-band polarization level is very small in the bulk area of \object{IRS5A}. And the $K$-band polarization found in east and south of \object{IRS5A} has no mid-IR counterparts; only the mid-IR feature in northeast of \object{IRS5A} coincides with $K$-band polarization.

Besides the two bright objects mentioned above, \object{B273A} and \object{IRS5C} are also visible in the VISIR $11.85\,\mu$m image; \object{IRS5C} is even brighter than \object{B273A} at this wavelength.

In addition, the mid-IR images at other wavelengths taken by TIMMI2 show only the two bright mid-IR objects, \object{UC1} and \object{IRS5A}, because of the lower angular resolution and sensitivity. Thus, these mid-IR images taken by TIMMI2 are not shown here, but only the flux densities of \object{UC1} and \object{IRS5A} are listed in Table~\ref{Tbl:Mir}.

\begin{table*}
\caption{Mid-IR flux densities of the YSOs in the \object{M\,17 UC1\,--\,IRS5} region}
 \centering
  \begin{tabular}{l c c c c c}    
\hline\hline
      &  $8.7\,\mu$m \tablefootmark{a}& $10.38\,\mu$m \tablefootmark{a} & $11.85\,\mu$m  & $17.72\,\mu$m \tablefootmark{a} & $20.6\,\mu$m \tablefootmark{c}  \\
      Object     & (mJy) & (mJy) & (mJy) & (mJy)  & (mJy) \\
   \hline 
   M\,17 B273A   & -  &  - &$46\pm22$ \tablefootmark{b}& - &- \\                                                                                        M\,17 UC1 &   $18700\pm1300$ & $7300\pm1000$ & $31300\pm1100$\tablefootmark{a} & $146700\pm29700$ & $128500\pm6800$ \\
    M\,17 IRS5A  & $3200\pm1600$ &  $6800\pm1500$ & $9700\pm1100$ \tablefootmark{a} & $130000\pm31000$ & $103900\pm5500$ \\
   M\,17 IRS5C  & - & - & $256\pm25$ \tablefootmark{b}& - & -\\

    \hline
  \end{tabular}
 \tablefoot{
\tablefoottext{a}{2\farcs0 aperture.}
 \tablefoottext{b}{0\farcs4 aperture.} \tablefoottext{c}{3\farcs2 aperture from K+02.}}
\label{Tbl:Mir}
\end{table*}   

\begin{figure}
\centering
\resizebox{\hsize}{!}{\includegraphics[width=0.48\textwidth]{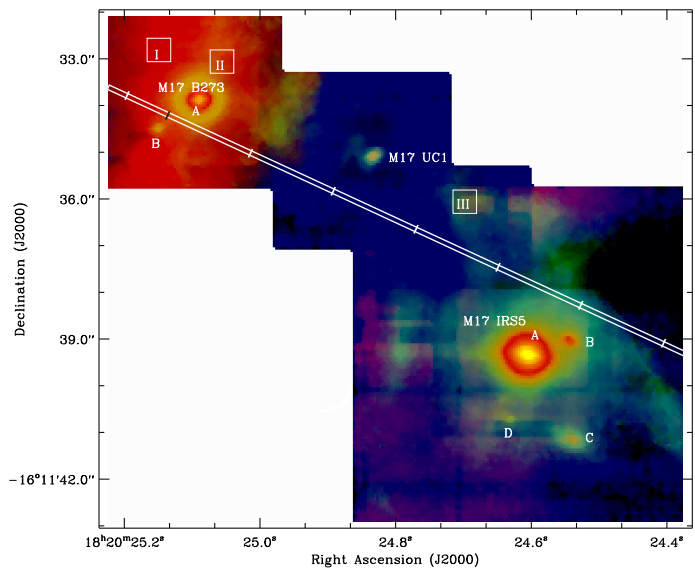}}
\caption{Three-color image created from the emission (line+continuum) of three near-IR lines (red: Br$\gamma$ $2.167\,\mu$m, green: H$_2$ $1-0$\,S(1) $2.122\,\mu$m, blue: \ion{He}{i} $2.059\,\mu$m). The three squares (I, II, and III) mark selected positions representing \ion{H}{ii} region, ionization front, and PDR. A strip line with tickmarks is also drawn perpendicular to the IF of M\,17 SW. Offset at each tickmark is $-1$, 0, $+2$, $+4$, $+6$, $+8$, $+10$, $+12$ in arcsec from northeast to southwest, respectively.}
\label{Fig:Lmp}
\end{figure}

\subsection{SINFONI near-IR spectroscopy of diffuse content}
\label{Sec:Sin}
The SINFONI observations are centered on \object{IRS5}, and are extended northeast to cover \object{UC1} and \object{B273}. Fig.~\ref{Fig:Lmp} shows the three-color composite of three lines. This figure reveals both the stellar content and diffuse emission in that area. For each object visible in Fig.~\ref{Fig:Lmp}, a SINFONI $H+K$ spectrum is available. However a spectral classification could only be obtained for \object{IRS5A} and \object{B273A} (Sect.~\ref{Sec:Spt}), while the S/N of the other spectra was not sufficient. Br$\gamma$ and $2.059\,\mu$m \ion{He}{i} $2^{1}$P$-$$2^1$S line emission are detected throughout the FOV. The location of the \ion{H}{ii} region and the IF is traced by the very strong Br$\gamma$ emission at the top-left corner. On the other hand, the PDR is characterized by $2.122\,\mu$m H$_2$ 1$-$0\,S(1) emission. We note H$_2$ emission surrounding \object{IRS5A} and H$_2$ emission coinciding with the bar-like IRN illuminated by \object{UC1}.

\subsubsection{Nebular emission lines}
\label{Sec:Nl}
\begin{figure*}
\centering
\resizebox{\hsize}{!}{\includegraphics[width=0.525\textheight,angle=90]{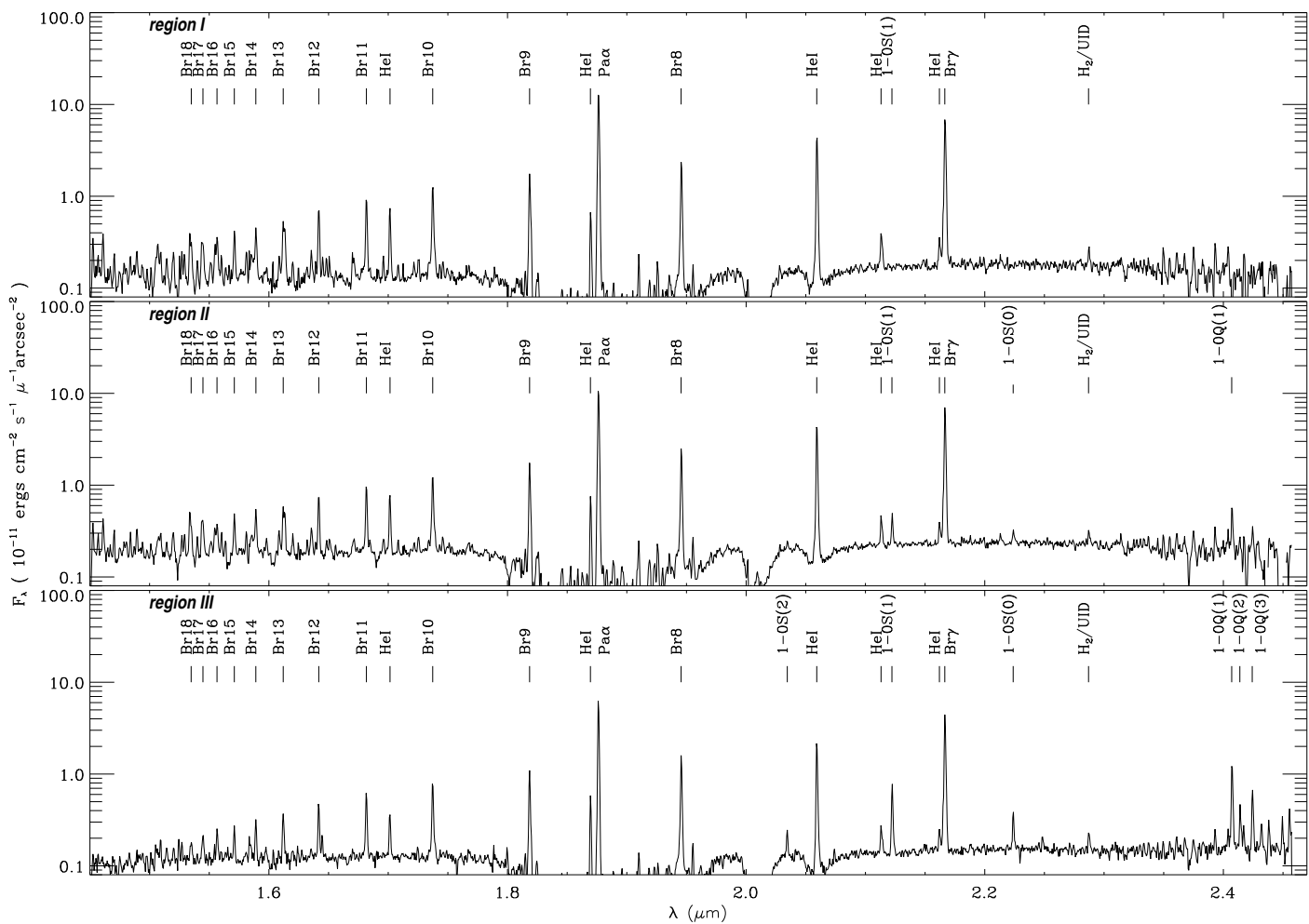}}
\caption{Ambient spectra of the three selected positions marked by squares in Fig.~\ref{Fig:Lmp}. The telluric feature in the range $1.8-2.1\,\mu$m is not corrected, because its strength is in proportion to the nebular continuum, which is much weaker than the emission lines. Thus the near-IR emission lines are little affected by the telluric feature.}
\label{Fig:Nl}
\end{figure*}

The spectra in the three selected regions are shown in Fig.~\ref{Fig:Nl}, and the observed lines are listed in Table~\ref{Tbl:Nl}. The hydrogen recombination lines such as Brackett series and Pa$\alpha$ are visible in all regions. Besides the prominent atomic hydrogen lines, five atomic helium lines are also observed; among them 2.059\,$\mu$m \ion{He}{i} is the strongest one. Molecular hydrogen emission lines longward $2\,\mu$m are also observed in region II and III. An emission feature at $2.287\,\mu$m is generally classified as H$_2$ $v=3-2$\,S(2) line  or unidentified (UID) line \citep[e.g.][]{2001MNRAS.328..419L}. However, the H$_2$/UID line might have been mismatched for the H$_2$ $v=3-2$\,S(2) line, because H$_2$ emission is not spatially coincident with this H$_2$/UID line in planetary nebulae \citep[e.g. NGC 7027,][]{2004PASJ...56..705O}. The most potential carrier of this UID line is [\ion{Se}{iv}] at 2.287\,$\mu$m, which is typically detected in highly excited ISM such as planetary nebulae and UC\ion{H}{ii} regions with hot O-type stars \citep[][and references therein]{2008AJ....135.1708B}. The ionizing sources (O4 binary) of M\,17 \ion{H}{ii} region are very likely to produce the 2.287\,$\mu$m \ion{Se}{iv} line in M\,17 SW. The 2.287\,$\mu$m UID line keeps roughly the same strength throughout the FOV of Fig.~\ref{Fig:Lmp}, indicating that this line traces the ionized gas region. Therefore, we attribute the 2.287\,$\mu$m line to [\ion{Se}{iv}].


\begin{table*}
\caption{Observed near-IR lines toward the \object{M\,17 UC1\,--\,IRS5} region}
 \centering
  \begin{tabular}{c c c c c c c}    
\hline\hline
    $\lambda$\,($\mu$m) & IDs  & Transition  & Eupper (K)\tablefootmark{a} & \multicolumn{3}{c}{$F\pm\delta F$($10^{-15}\,\mathrm{erg}\,\mathrm{cm}^{-2}\,\mathrm{s}^{-1}\,\mathrm{arcsec}^{-2}$)}\\
\hline
Obs. & & &  & region I & region II & region III \\
  
   \hline

    1.5349 &       H I &                   18$-$4    &     &       4.3 $\pm$       2.0 &       4.9 $\pm$       2.1 &       1.1 $\pm$       0.4   \\
    1.5447 &       H I &                    17$-$4   &     &       4.2 $\pm$       1.5 &       4.8 $\pm$       1.6 &       1.7 $\pm$       0.3   \\
    1.5566 &       H I &                    16$-$4   &     &       3.7 $\pm$       1.2 &       2.6 $\pm$       1.0 &       1.8 $\pm$       0.6   \\
    1.5710 &       H I &                    15$-$4   &     &       3.3 $\pm$       1.1 &       4.9 $\pm$       2.2 &       2.1 $\pm$       0.2   \\
    1.5891 &       H I &                    14$-$4   &     &       5.4 $\pm$       0.6 &       5.7 $\pm$       0.7 &       2.1 $\pm$       0.6   \\
    1.6120 &       H I &                    13$-$4   &     &       8.4 $\pm$       1.9 &       7.6 $\pm$       2.6 &       3.9 $\pm$       1.2   \\
    1.6417 &       H I &                    12$-$4   &     &       6.8 $\pm$       2.1 &       5.6 $\pm$       2.6 &       4.5 $\pm$       0.4   \\
    1.6817 &       H I &                    11$-$4   &     &      11.6 $\pm$       0.8 &      11.4 $\pm$       0.5 &       7.0 $\pm$       0.5   \\
    1.7013 &      He I &          4$^3$D$-$3$^3$P    &     &       7.3 $\pm$       1.6 &       7.4 $\pm$       1.6 &       3.3 $\pm$       0.3   \\
    1.7372 &       H I &                    10$-$4   &     &      17.6 $\pm$       1.1 &      16.8 $\pm$       1.0 &       9.9 $\pm$       0.6   \\
    1.8185 &       H I &                    9$-$4    &     &      23.3 $\pm$       1.3 &      21.8 $\pm$       1.8 &      13.0 $\pm$       1.3   \\
    1.8696 &      He I &          4$^3$F$-$3$^3$D    &     &       7.9 $\pm$       0.9 &       7.8 $\pm$       0.8 &       5.7 $\pm$       0.7   \\
    1.8762 &       H I &                    4$-$3    &     &     183.9 $\pm$       1.0 &     147.8 $\pm$       1.1 &      74.8 $\pm$       0.7   \\
    1.9457 &       H I &                    8$-$4    &     &      31.7 $\pm$       1.0 &      32.1 $\pm$       1.5 &      19.5 $\pm$       0.8   \\
    2.0344 &     H$_2$ &            $v=$1$-$0 S(2)   &  7584   &   --  &       0.6 $\pm$       0.4 &       1.8 $\pm$       0.4   \\
    2.0592 &      He I &          2$^1$P$-$2$^1$S    &     &      61.2 $\pm$       0.6 &      60.8 $\pm$       0.6 &      28.4 $\pm$       0.3   \\
    2.1132 &      He I &    4$^{1,3}$$-$3$^{1,3}$P   &     &       4.2 $\pm$       0.8 &       4.0 $\pm$       0.3 &       2.4 $\pm$       0.4   \\
    2.1224 &     H$_2$ &            $v=$1$-$0 S(1)   &  6956   &  --  &       3.3 $\pm$       0.4 &       8.5 $\pm$       0.2   \\
    2.1620 &      He I &          7$^3$F$-$4$^3$D    &     &       3.1 $\pm$       0.4 &       3.0 $\pm$       0.5 &       1.5 $\pm$       0.4   \\
    2.1667 &       H I &                     7$-$4   &     &      99.4 $\pm$       0.7 &      92.8 $\pm$       0.4 &      57.3 $\pm$       0.3   \\
    2.2239 &     H$_2$ &             $v=$1$-$0 S(0)  &  6471   &       0.0 $\pm$       0.7 &       1.5 $\pm$       0.5 &       3.2 $\pm$       0.6   \\
    2.2873 & [\ion{Se}{iv}] &  $^2$P$_{3/2}-^2$P$_{1/2}$                           &     &       1.2 $\pm$       0.4 &       1.4 $\pm$       0.4 &       1.5 $\pm$       0.5   \\
    2.4073 &     H$_2$ &           $v=$1$-$0 Q(1)    &  6149   &   --    &       7.2 $\pm$       1.0 &      15.7 $\pm$       0.5   \\
    2.4141 &     H$_2$ &            $v=$1$-$0 Q(2)   &  6471   &   --   &    --   &       4.8 $\pm$       0.4   \\
    2.4244 &     H$_2$ &           $v=$1$-$0 Q(3)    &  6956  &    --  &       4.1 $\pm$       1.2 &       6.8 $\pm$       0.7   

\\
    \hline
  \end{tabular}
\tablefoot{Lines detected in the averaged spectra of three selected regions. The flux of each line is averaged over each region. \\
\tablefoottext{a}{Excitation energy adopted from \citet{1984CaJPh..62.1639D}.}}
\label{Tbl:Nl}
\end{table*}

\subsubsection{$\ion{H}{i}/\mathrm{H}_2$ transition zone of M\,17 SW}
\label{Sec:HH}
\begin{figure}
\centering
\resizebox{\hsize}{!}{\includegraphics[angle=90,width=0.48\textwidth]{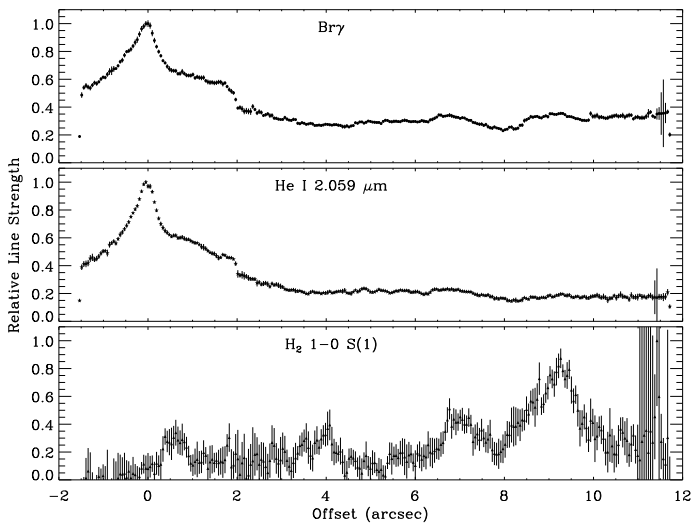}}
\caption{Line intensity variations of three representative lines -- Br$\gamma$, 2.059\,$\mu$m \ion{He}{i}, and H$_2$ $1-0$\,S(1) -- along the strip line which is perpendicular to the IF of M\,17 SW (outlined in Fig.~\ref{Fig:Lmp}). The zero-point is defined at $18\colon20\colon25.136$, $-16\colon11\colon34.21$ (J2000), at the maximum strength of the IF; negative offsets point toward the \ion{H}{ii} region, while positive offsets point toward the PDR. }
\label{Fig:2d}
\end{figure}

The coexistence of Br$\gamma$ and H$_2$ $1-0$\,S(1) in region II and III implies that region II and III maybe part of the \ion{H}{i}/H$_2$ transition zone of \object{M\,17 SW}. In order to investigate the scale size of this 
$\ion{H}{i}/\mathrm{H}_2$ transition zone, Fig.~\ref{Fig:2d} shows line strength variations of Br$\gamma$, 2.059\,$\mu$m \ion{He}{i} and H$_2$ $1-0$\,S(1) along the strip line denoted in Fig.~\ref{Fig:Lmp}. Peaks of Br$\gamma$ and 2.059\,$\mu$m \ion{He}{i} both occur around IF. Specifically, the 2.059\,$\mu$m \ion{He}{i} peak occurs $\sim0\farcs1$ closer to the \ion{H}{ii} region than the Br$\gamma$ peak, which defines the IF. Molecular hydrogen starts to appear at offset $\sim+0\farcs4$, closer to the PDR. Along the strip line, the line strengths of Br$\gamma$ and 2.059\,$\mu$m \ion{He}{i} show almost identical variations, which both experience fast growth until the IF, rapid drop (offset $<+0\farcs
5$), shallow decline (offset between $+0\farcs5-+2\arcsec$), and roughly constant baseline (offset longwards $+2\arcsec$). The H$_2$ $1-0$\,S(1) line strength has another type of variation, which shows four peaks along the strip line. The coexistence of atomic gas (\ion{H}{i}, \ion{He}{i}) and molecular gas (H$_2$) in the range $+0.4\arcsec-+11.7\arcsec$ and the trend further into the PDR indicates that the 
$\ion{H}{i}/\mathrm{H}_2$ transition zone of M\,17 SW almost starts from the IF and extends toward the cloud core with a projected scale size more than $12\arcsec$ (0.12\,pc). However, one has to be cautious because the emission from the HII region and the PDR may partly overlap along the LOS if the configuration is not strictly seen edge-on; the \ion{H}{ii} region emission lines originate from the surface of the PDR, while PDR emission lines form inside the PDR. In such a case, the co-existence of emission lines like in Fig.~\ref{Fig:2d} will not tell us the true story about $\ion{H}{i}/\mathrm{H}_2$ transition zone. Comparison with a strictly edge-on PDR will help to clarify whether this scale size of $\ion{H}{i}/\mathrm{H}_2$ transition zone in \object{M\,17 SW} is reliable or not. 

The Orion Bar is a dense PDR like \object{M\,17 SW}, but nearly edge-on \citep[and references therein]{2005ApJ...630..368A}. We measured the scale size of $\ion{H}{i}/\mathrm{H}_2$ transition zone in the Orion Bar according to Fig.~3 of \citet{1985MNRAS.215P..31H} to be around 20\arcsec, which is consistent with the distributions of H$_2$ rotational lines \citep{2005ApJ...630..368A}. Due to the edge-on configuration of the Orion Bar, the scale size measured above is identical to the real scale zie of the $\ion{H}{i}/\mathrm{H}_2$ transition zone, i.e. 0.044\,pc at the distance of Orion \citep[450 pc;][]{2000ApJ...544L.133H}. However, the $\ion{H}{i}/\mathrm{H}_2$ transition zone in \object{M\,17 SW} is at least three times Orion Bar in size, which can not be simply explained by the differential properties between the two PDRs. We speculate that the presence of the three emission lines in Fig.~\ref{Fig:2d} is simply a result of geometric projection. Unlike the Orion Bar, \object{M\,17 SW} is indeed inclined from the LOS with a substantial angle. This result confirms the conclusion of a recent study about the LOS structure of \object{M\,17 SW} based on H$_2$ rotational emission \citep{2013ApJ...774L..14S}.

\subsubsection{H$_2$ excitation}
\label{Sec:H2}
As seen in Table~\ref{Tbl:Nl}, six H$_2$ emission lines are detected. The two most prominent H$_2$ lines are H$_2$ $1-0$\,Q(1) at 2.407\,$\mu$m and H$_2$\,$1-0$\,S(1) at 2.122\,$\mu$m. H$_2$ emission lines in PDR are generally thought to have two physical mechanisms\,--\,fluorescent excitation by FUV photons and thermal excitation in shock fronts. Practically, near-IR H$_2$ fluorescent spectra have been observed for a variety of classical PDRs such as associated with \object{NGC\,2023}, Orion Bar, and the northern bar in \object{M\,17} \citep{1987ApJ...318L..73G,1985MNRAS.215P..31H,1989ApJ...336..207T}. Meanwhile, a variety of H$_2$ spectra have been found to be produced in the shock fronts associated with the jets/outflows found in PDRs \citep[e.g.][]{2007A&A...465..931N,2008A&A...489..229M,2010ApJ...713..883B}. Although these two mechanisms both act occasionally for H$_2$ emission found in PDRs, they can be distinguished since the corresponding H$_2$ line ratios are different.

\begin{figure*}
\centering
\resizebox{\hsize}{!}{\includegraphics{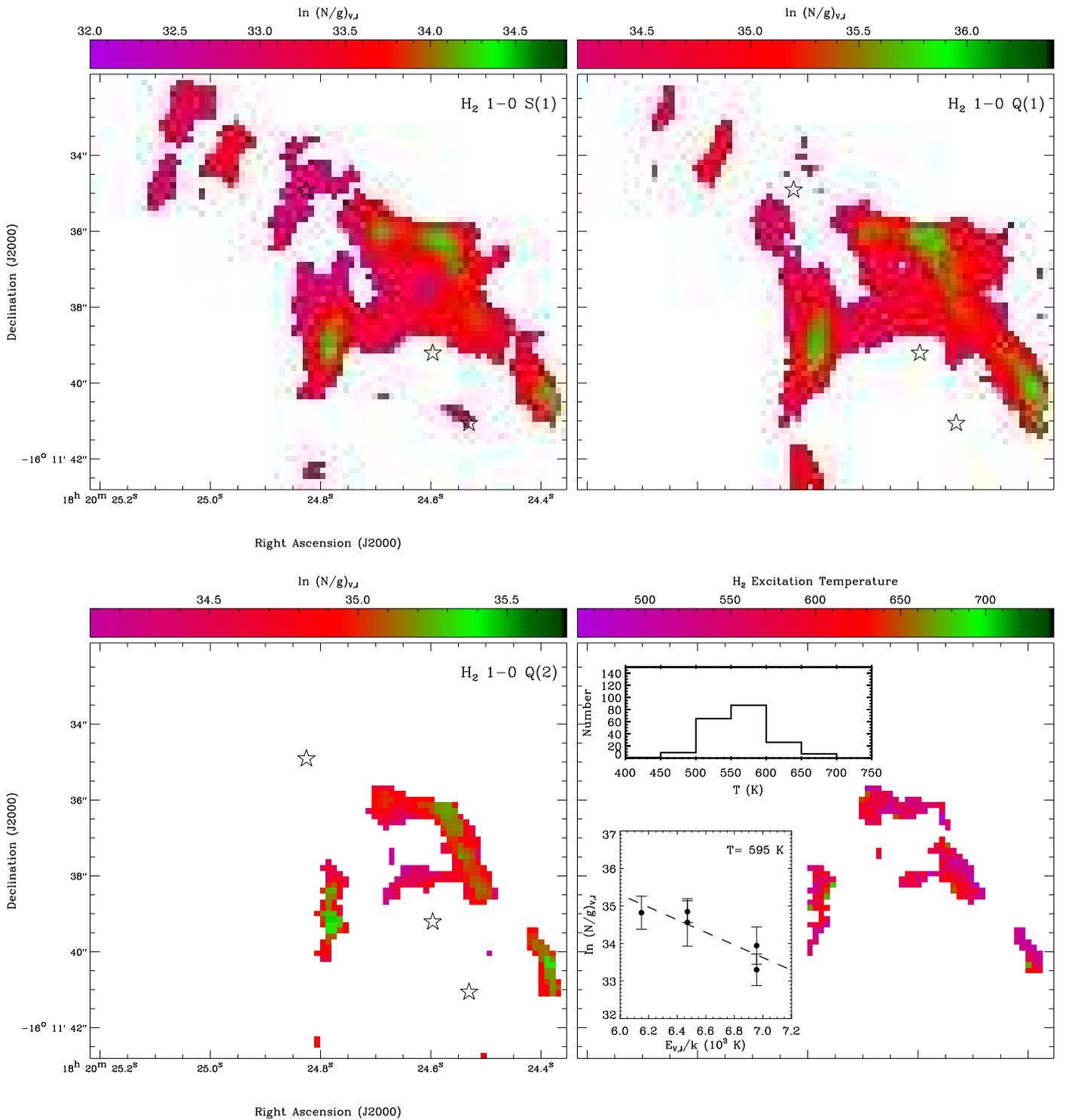}}
\caption{Column densities of H$_2$ 1$-$0\,S(1), 1$-$0\,Q(1) and 1$-$0\,Q(2) (quadrants I to III) as well as the H$_2$ excitation temperature map (quadrant IV). In quadrants I to III, \object{IRS5A,C} and \object{UC1}'s positions are marked by asterisks. In quadrant IV, the ro-vibrational diagram based on the H$_2$ lines (H$_2$ 1$-$0\,S(1), 1$-$0\,S(0), 1$-$0\,Q(1), 1$-$0\,Q(2), and 1$-$0\,Q(3)) averaged over all collected pixels is plotted, as well as the histogram density of H$_2$ excitation temperature. }
\label{Fig:H2ext}
\end{figure*}

A common way of characterizing the H$_2$ emission is to evaluate the gas temperature in the framework of a ro-vibrational diagram, which is a plot of the observed column density against the energy of the upper level. The column density, $N_{j}$, of the upper level of a given transition can be calculated from the measured line intensity, $I$, of the corresponding H$_2$ line via the following formula: $$N_j=\frac{4\,\pi \,\lambda_{j}\,I}{A_{j}\,h\,c},$$ where $\lambda_j$ (the rest wavelength) and $A_j$ (the Einstein A-coefficient) are taken from \citet{1977ApJS...35..281T}. If collisional deexcitation plays dominant role, the H$_2$ molecule will be in LTE and the energy level obeys the Boltzmann distribution. In such distribution, the relative column densities of any two excitation levels can be expressed in terms of excitation temperature $T_{\mathrm{ex}}$:
$$\frac{N_i}{N_j}=\frac{g_i}{g_j}\,\mathrm{exp}\left[\frac{-(E_i-E_j)}{k\,T_{\mathrm{ex}}}\right],$$ where $g_j$ is the degeneracy, $E_j$ is the excitation energy taken from \citet{1984CaJPh..62.1639D}, and $k$ is the Boltzmann constant. If the gas is thermalized at a single temperature, the plot of the logarithm of $N_{j}$ to $g_{j}$ ratio against the level energy will reveal a straight line, whose slope provides the reciprocal of the excitation temperature.

We implied this method to evaluate the excitation temperature for the detected H$_2$ emission lines, assuming the same dust attenuation for all lines. Fig.~\ref{Fig:H2ext} shows the extinction-uncorrected column densities of H$_2$ 1$-$0\,S(1), H$_2$ 1$-$0\,Q(1), and H$_2$ 1$-$0\,Q(2). We excluded the H$_2$ 1$-$0\,S(2) line in plotting the ro-vibrational diagram, because this line is too weak to be significantly detected in most FOV. Except H$_2$ 1$-$0\,S(1) and 1$-$0\,Q(1), the remaining three H$_2$ lines are weak, although stronger than 1$-$0\,S(2). To increase the $S/N$ ratios for these weak H$_2$ lines, we degraded the angular resolution of the SINFONI data by rebinning the data with $3\times3$ array. The resulting $S/N$ ratios are increased by a factor of $\sim3$. Note that rebinning is applied merely when the H$_2$ emission lines are analyzed. Thus, only Fig.~\ref{Fig:H2ext} and Fig.~\ref{Fig:Qsr} are plotted in the rebinned pixel scale. For H$_2$ 1$-$0\,S(1) and 1$-$0\,Q(1), a threshold of $S/N\geqslant3$ is used; for the remaining three H$_2$ lines, $S/N\geqslant2$ is applied. The lower-right panel in Fig.~\ref{Fig:H2ext} presents the map of $T_\mathrm{ex}$ for H$_2$ emission, and the histogram of excitation temperatures, as well as the ro-vibirational diagram based on the column densities averaged throughout the collected pixels. The ro-vibrational diagram shows that all five H$_2$ lines lie along a straight line whose slope corresponds to an excitation temperature $\approx600\,\mathrm{K}$. We note that the determined $T_\mathrm{K}$ for each rebinned pixel might have large errors because the adopted H$_2$ lines just cover a very narrow range of excitation temperature ($6100-7000\,\mathrm{K}$). Nevertheless, the statistical $T_\mathrm{ex}$ of the H$_2$ gas is more meaningful than a specific value for each rebinned pxiel. With the histogram of $T_\mathrm{ex}$ for all rebinned pixels, we found a peak around 575\,K. A single excitation temperature for the H$_2$ emission indicates that the energy levels are thermally excited. Other discriminators for the two excitation mechanisms of H$_2$ emission such as H$_2$ 1$-$0\,S(1)/2$-$1\,S(1) ratio and H$_2$ ortho-to-para ratio both indicate thermal excitation.

For fluorescence emission, each vibrational level has $N_j/g_j$ lying along a separate ``branch'', and the rotational population within each level can be approximated by a thermal distribution. If there are transitions from several levels, a curved line in the ro-vibrational diagram would therefore provide evidence for non-LTE process \citep{2008A&A...489..229M}. The H$_2$ lines at higher levels, e.g. $v=3-2$, can be useful to distinguish between the two mechanisms. In the case of the \object{M\,17 UC1\,--\,IRS5} region, the merely available $v=1-0$ transitions cannot be used to discriminate between shock and fluorescent excitation, because they will lie along a straight line in the ro-vibrational diagram for both mechanisms if uncertainties are considered. Moreover, in dense PDRs the lower H$_2$ levels will be thermalized as like in shock fronts. This degeneracy between dense PDRs and shock fronts prevents us to exactly characterize the H$_2$ emission from the ro-vibrational diagram. This approach is more meaningful for the regions with moderate/strong H$_2$ emission at all transitions. Indeed, it is curious to know whether weak H$_2$ emission area possessing the same excitation mechanism with the stronger H$_2$ emission area.

\begin{figure}
\centering
\includegraphics[width=0.48\textwidth]{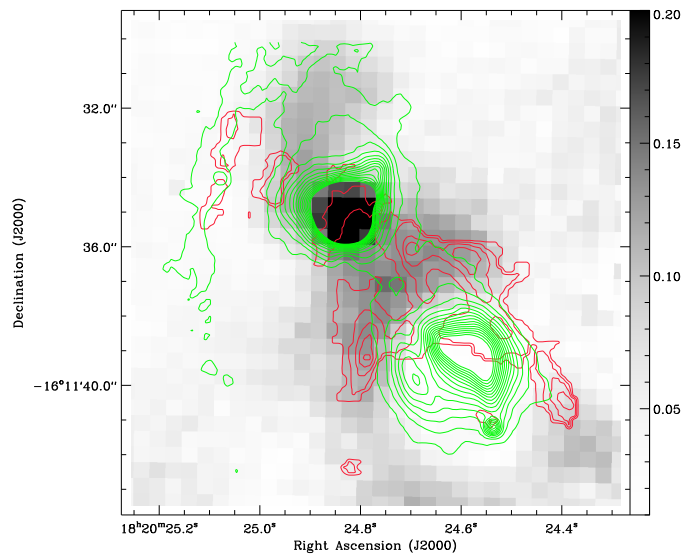}
\caption{H$_2$ 1$-$0\,S(1) line flux (red contours) overlaid on the IRSF/$K$ polarization degree image (grey scale; CZ+12). H$_2$ 1$-$0\,S(1) contour levels start from  $2\times10^{-15}\,\mathrm{erg}\,\mathrm{cm}^{-2}\,\mathrm{s}^{-1}\,\mathrm{arcsec}^{-2}$ with four intervals, each of $2\times10^{-15}\,\mathrm{erg}\,\mathrm{cm}^{-2}\,\mathrm{s}^{-1}\,\mathrm{arcsec}^{-2}$. Green contours correspond to the VISIR $11.85\,\mu$m emission. }
\label{Fig:PolH2}
\end{figure}

\begin{figure}
\centering
\includegraphics[width=0.48\textwidth]{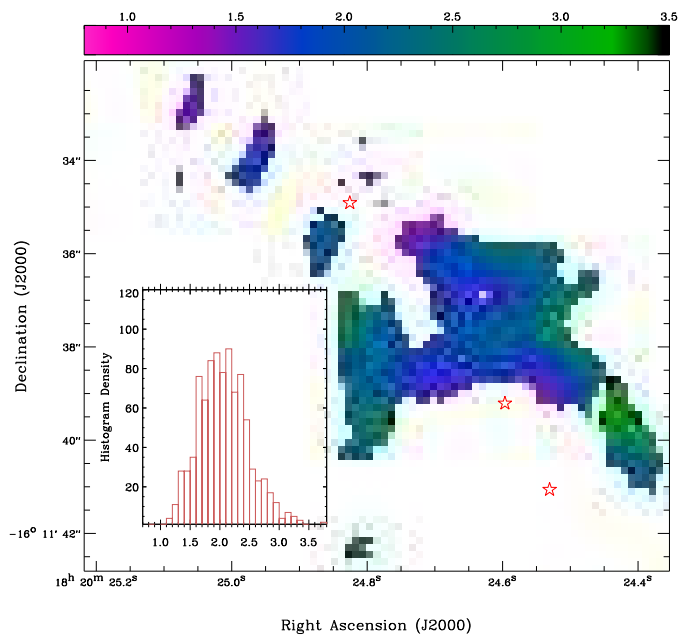}
\caption{Map of H$_2$ 1$-$0\,Q(1)/1$-$0\,S(1) ratio, and corresponding histogram density at lower left corner. The three asterisks are as same as in Fig.~\ref{Fig:H2ext}.}
\label{Fig:Qsr}
\end{figure}

In an \ion{H}{ii} region, another attempt to qualitatively characterize the H$_2$ emission comes from the comparison of molecular and atomic hydrogen emission. The molecular-to-atomic line ratio, particularly the ratio H$_2$ 1$-$0\,S(1)/Br$\gamma$, is $\leqslant1$ in active galaxies (starburst, Seyfert, and ultra-luminous infrared galaxies) and star-forming regions \cite[][and reference therein]{2005MNRAS.358..765H}. Conversely, this ratio is $\geqslant1$ in outflow regions \citep{2005MNRAS.358..765H, 2008A&A...489..229M}. All regions showing H$_2$ 1$-$0 S(1) emission yield H$_2$ 1$-$0\,S(1)/Br$\gamma\ll1$. The extremely strong hydrogen atomic emission observed here indicates very intense incident stellar radiation field, which makes the FUV pumping as the most likely excitation mechanism for the molecular hydrogen emission observed in our study. Further indirect evidence also points out that FUV fluorescence is the most plausible excitation mechanism. For instance, H$_2$ 1$-$0\,S(1) and H$_2$ 1$-$0\,Q(1) emission seen in the top two panels in Fig.~\ref{Fig:H2ext} shows some filament structures parallel to the IF. This H$_2$ emission can not be characterized by the aforementioned approaches, since just emission of 1$-$0 S(1) and 1$-$0 Q(1) are available. One may argue, however, that the H$_2$ emission lying parallel to IF can be produced in shock waves driven by the IF. An identical case is the Orion Bar PDR which shows H$_2$ emission behind itself with a ratio of H$_2$ 1$-$0\,S(1)/2$-1$\,S(1) $\sim3$ resembling shock excitation \citep{1985MNRAS.215P..31H}. Nevertheless, \citet{1990ApJ...352..625B} proposed a high-density PDR model to explain the observed line ratio because the typical shock speed, $\leqslant3\,\mathrm{km\,s}^{-1}$, of an expanding \ion{H}{ii} region driving into molecular gas is too low to significantly excite low-$v$ H$_2$ transitions. Considering the common properties between M\,17 SW and Orion Bar, we tend to neglect shock excitation for the IF-parallel H$_2$ emission. Complementally, in Fig.~\ref{Fig:PolH2} the IRN illuminated by \object{IRS5A} and by \object{UC1} (the one discussed in Sect.~\ref{Sec:En}) spatially coincides with H$_2$ emission. H$_2$ fluorescent emission has been found for some large-scale IRN associated with star-forming regions such as \object{NGC\,2023} and Orion Bar \citep{1987ApJ...318L..73G,1990ApJ...352..625B}. Although the IRN here are associated with luminous IR stars, we speculate that this coincidence is an indirect evidence in favor of FUV fluorescence excitation.

In Fig.~\ref{Fig:Qsr} we show the H$_2$ 1$-$0\,Q(1)/1$-$0\,S(1) ratio map, which covers most of the H$_2$ emitting regions. The H$_2$ 1$-$0\,Q(1)/1$-$0\,S(1) ratio intrinsically varies with the physical conditions of H$_2$ gas. \citet{1998ApJ...499..799L} modeled H$_2$ line ratios detected in Orion Bar and Orion S including H$_2$ 1$-$0\,Q(1)/1$-$0\,S(1) ratio for several set of PDR models and thermal excitation by shock front as well. From their modeling, we find distinct H$_2$ 1$-$0\,Q(1)/1$-$0\,S(1) ratios between PDR models and shocked thermal excitation \citep[see Table~1 in][]{1998ApJ...499..799L}. The H$_2$ 1$-$0\,Q(1)/1$-$0\,S(1) ratios of the H$_2$ emission in M\,17 SW are mostly in the range $1.0-3.0$. The bulk of the H$_2$ emission regions have a H$_2$ 1$-$0\,Q(1)/1$-$0\,S(1) ratio $\sim1.8$, which is between the ratios of two PDR models of $n_\mathrm{H}=10^6\,\mathrm{cm}^{-3}$, $G_0=10^4$, $T_0=500\,\mathrm{K}$, and $n_\mathrm{H}=10^6\,\mathrm{cm}^{-3}$, $G_0=10^5$, $T_0=1000\,\mathrm{K}$, respectively. In contrast, the H$_2$ 1$-$0\,Q(1)/1$-$0\,S(1) ratio predicted by the shocked thermal model is 0.7, much lower than the ratios observed here. The extinction-corrected H$_2$ 1$-$0\,Q(1)/1$-$0\,S(1) ratio will be smaller than the observed one, however, the decrease caused by dereddenning is not big enough to lower this ratio to a thermal value.


We note quite uniform distributions of line ratios such as H$_2$ 1$-$0\,S(1)/2$-$1\,S(1), H$_2$ 1$-$0\,S(1)/Br$\gamma$, and H$_2$ 1$-$0\,Q(1)/1$-$0\,S(1) across all H$_2$ emitting areas in the \object{M\,17 UC1\,--\,IRS5} region, which indicates a uniform mechanism for the entire H$_2$ emission. All the above properties of H$_2$ emission can be well explained in the scenario that the H$_2$ molecules inside M\,17 SW are initially pumped to vibrational states by FUV fluorescence, and then are deexcitated from high-$v$ levels to low-$v$ levels due to the high-frequency collision inside the high-density PDR. Due to this process, H$_2$ emission is mostly in the form of low-$v$ vibrational states and pure rotational states. Besides the $v=1-0$ H$_2$ emission lines reported here, \citet{2013ApJ...774L..14S} reported mid-IR pure-rotation H$_2$ emission at $v=0-0$ states toward \object{M\,17 SW}.

\begin{figure*}
\centering
\includegraphics[width=0.905\textwidth]{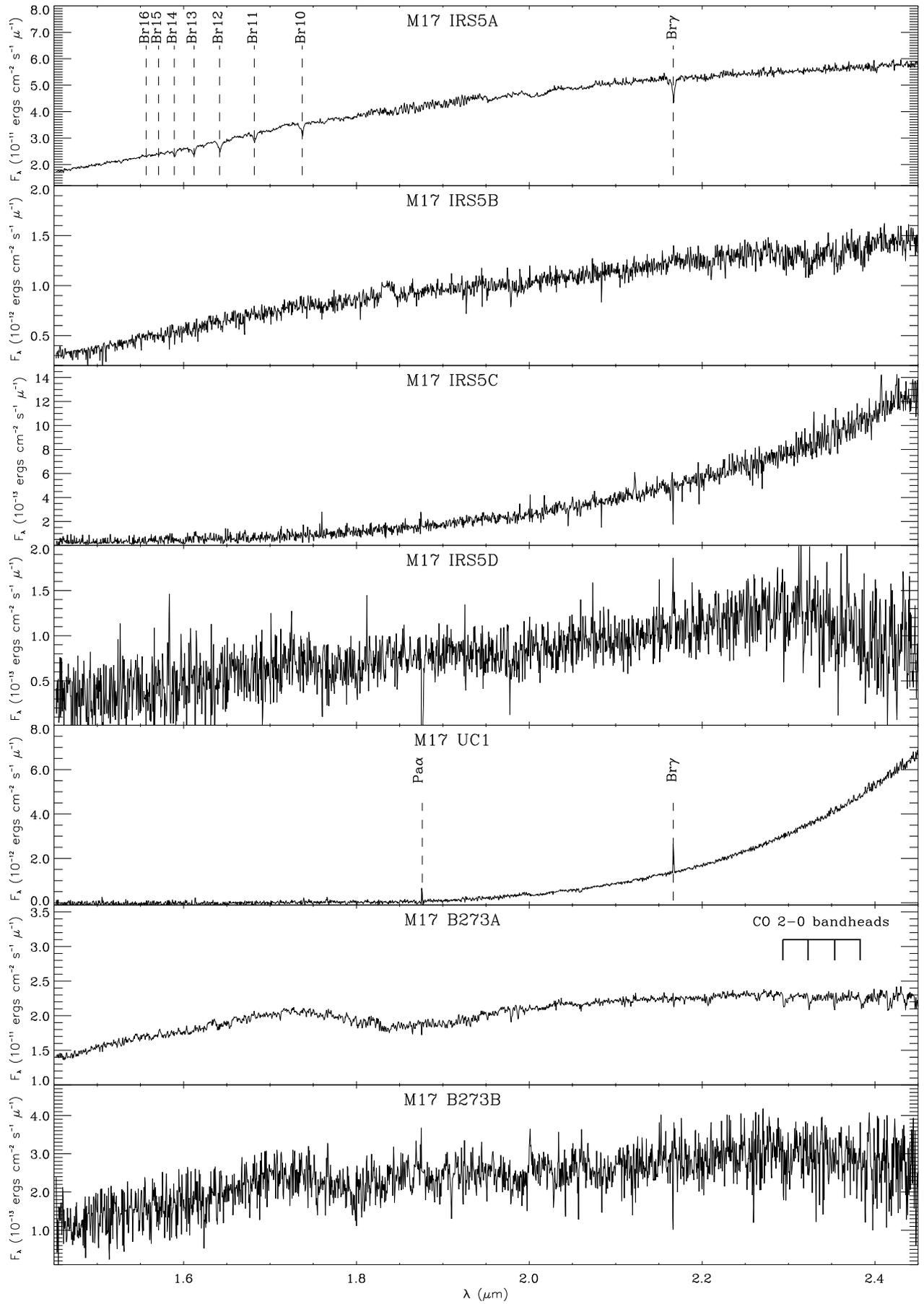}
\caption{Flux-calibrated SINFONI $H+K$ spectra of the point sources in M\,17\,UC1--IRS5 region.  All natural spectral features are marked, e.g. \ion{H}{i} emission/absorption lines and CO $2-0$ bandheads in absorption.}

\label{Fig:Af}
\end{figure*}


\begin{figure*}
\centering
\includegraphics[width=0.35\textwidth,angle=90]{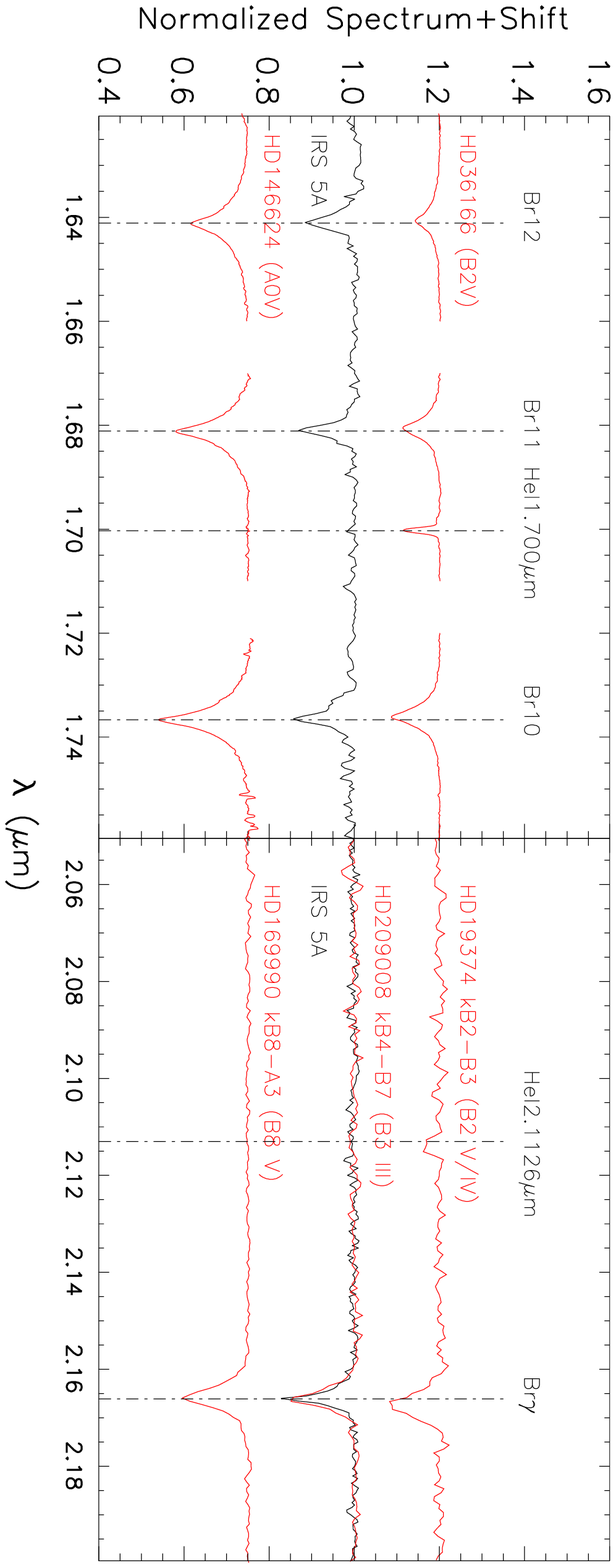}
\caption{Normalized SINFONI $H+K$ spectrum of IRS5A (black curve). For comparison, the spectra of several reference stars are displayed; they are adopted from the literature as follows: B2\,V (\object{HD\,36166}), A0\,V (\object{HD\,146624}) from \citet{2005ApJS..161..154H}; B2\,V/IV (\object{HD\,19374}), B3\,III (\object{HD\,209008}), B8\,V (\object{HD\,169990}) from \citet{1996ApJS..107..281H}. All reference spectra are converted to a resolution identical to SINFONI's resolving power.}
\label{Fig:Spc1}
\end{figure*}

\subsection{SINFONI near-IR spectroscopy of point sources \label{Sec:Spt}}
The SINFONI $H+K$ spectra of the YSOs in M\,17 UC1\,--\,IRS5 region, are crucial to understand their fundamental properties such as effective temperature and luminosity class via analyzing the characteristic spectral features (see Fig.~\ref{Fig:Af}). Objects located inside an HII region or a PDR require special care due to the contamination by nebular emission. The nebular contamination is evaluated in an annulus closely encircling the stellar contours. The spectra of \object{IRS\,5A} and \object{B273A} have sufficient $S/N$ to maintain a reliable classification, while the other sources are too faint. 

The most prominent lines in Fig.~\ref{Fig:Af} are the hydrogen Brackett absorption lines seen in \object{IRS5A}, the CO 2$-$0 bandhead absorption for \object{B273A}, and the \ion{H}{i} Pa$\alpha$ and Br$\gamma$ emission for \object{UC1}. The \ion{H}{i} emission lines of \object{UC1} are reminiscent of the common emission features of YSOs, which are due to the accretion flows falling onto the YSOs. However, we note from Fig.~\ref{Fig:Lmp} that the emission lines mostly form in the two reflection lobes of \object{UC1}, and very little part directly comes from \object{UC1} which is mostly obscured by its edge-on disk. The \ion{H}{i} emission lines seen in \object{UC1}'s spectrum arise most likely in the associated HCHII region. The shape of \ion{H}{i} emission region oriented perpendicularly to the edge-on circumstellar disk of \object{UC1}, is suggested to trace the relevant HCHII region expanding preferentially along the polar direction because of the lower-density gas in polar regions.



\begin{figure}
\centering
\resizebox{\hsize}{!}{\includegraphics[angle=90]{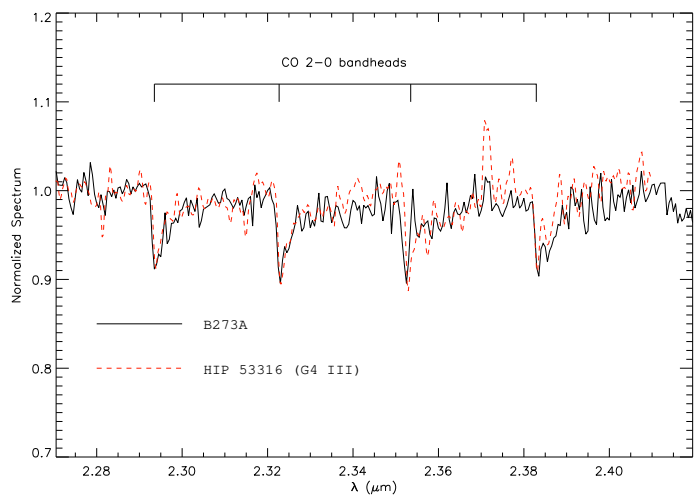}}
\caption{Normalized SINFONI spectrum of B273A (black) in the wavelength range $2.27-2.42\,\mu$m, covering CO $2-0$ handhead absorption. For comparison, the reference spectrum of mid-G giant \object{HIP\,53316} \citep[SpT G4\,III;][]{1997ApJS..111..445W} is shown in red. The reference spectrum ($R\sim3000$) has been smoothed to the spectral resolution of SINFONI spectrum.}
\label{Fig:Spc2}
\end{figure}

The normalized spectra of \object{IRS\,5A} and \object{273A} are shown in Fig.~\ref{Fig:Spc1} and \ref{Fig:Spc2}, respectively. \object{IRS5A}'s spectrum displays strong hydrogen Brackett absorption lines. The sole characteristic line in the $K$-band, Br$\gamma$, indicates a temperature class later than B3, since its effective temperature is not high enough for the growth of 2.113\,$\mu$m \ion{He}{i} line which appears for early-B stars\,\citep{1996ApJS..107..281H}. The Br$\gamma$ equivalent width\,(EW) is quantified to disentangle the ambiguity between mid-B\,(kB4--B7) and late-B/early-A\,(kB8--A3); the latter shows Br$\gamma$ EW greater than 8\,{\AA} while the former shows Br$\gamma$ EW between 4\,{\AA} and 8\,{\AA}\,\citep{1996ApJS..107..281H}. The Br$\gamma$ EW of \object{IRS5A} is measured to be $4.8\,\AA$, constraining its temperature class to be kB4--B7. How does the $K$-band temperature class link to the optical spectral class? In the study by \citet{1996ApJS..107..281H}, six reference stars with known optical spectral type are assigned to a $K$-band spectral type of kB4--B7. Four out of the six stars have optical spectral classes of B3--B6, with three dwarfs and one giant. The remaining two are optically classified late-B/early-A supergiants. \citet{1996ApJS..107..281H} stressed that the spectral types solely based on $K$-band spectrum are not sensitive to surface gravity and show ambiguity between mid-B stars and late-B/early-A supergiants.  

The $H$-band spectrum provides more spectral type indicators. The presence of strong Br$11$ line and weak 1.701\,$\mu$m\,\ion{He}{i} line (EW\,$<0.2$\,{\AA}) both point out a spectral class later than early-B \citep{1998AJ....116.1915H}. Moreover, the Br$11$ EW of \object{IRS5A} is determined as $4.1\,\AA$, suggesting mid-B spectral class. With the known spectral type range of \object{IRS5A}, it is possible to distinguish dwarfs and supergiants based on the Br$11$/1.701\,$\mu$m\,\ion{He}{i} ratio. \citet{1998AJ....116.1915H} found for early/mid-B stars that this ratio is consistently larger for dwarfs than for supergiants of the same spectral class; e.g. this ratio is about 3.0 in average for B4--B7 supergiants. In the case of \object{IRS5A}, the large ratio of Br$11$ to 1.701\,$\mu$m\,\ion{He}{i} ($\gtrsim20$) indicates a luminosity class close to dwarf.

Some \ion{He}{i} lines (e.g. 1.701\,$\mu$m, 2.113\,$\mu$m) are crucial diagnostic lines to separate early-B stars from later type stars \citep[e.g.][]{2005ApJS..161..154H, 2005A&A...440..121B}. Moreover, the difficulty in distinguishing dwarfs and giants solely based on near-IR spectral features also prevent us to better constrain the luminosity class of \object{IRS5A}. According to the $H+K$ spectral characteristics, we suggest a spectral type of B3--B7\,$\mathrm{V}/\mathrm{III}$ for \object{IRS5A}.

\object{B273A}'s $H$-band spectrum does not show any significant characteristic line. Its $K$-band spectrum only shows 2.3\,$\mu$m CO $2-0$ bandhead absorption. The CO bandhead absorption longwards of 2.29\,$\mu$m is typical of late-type stars \citep[e.g.][]{1997ApJS..111..445W}, whose outer atmospheric layers have the proper temperatures ($\sim1000-3000$\,K) to produce such features. Moreover, low-mass Class II/III YSOs (temperatures identical to those of late-type stars) might also show CO bandhead in absorption despite of the circumstellar material because the circumstellar veiling is less than that of Class I YSOs and consequently overwhelmed by the photospheric feature \citep{1996A&A...306..427C}. 

The CO bandhead absorption is found to be tightly related with the effective temperature and luminosity class for giants and supergiants. The EW of CO bandhead increases linearly with temperature declining for giants and supergiants, respectively, and the later has higher EW than the former of the same effective temperature \citep[e.g.][]{2012A&A...539A.100G}. The CO bandhead of \object{B273A} coincides in depth and width with that of the reference star, a G4 giant (see Fig.~\ref{Fig:Spc2}). Indeed, a supergiant needs to have earlier temperature class to match the EW of a giant's CO bandhead. We measured the EW of CO bandhead for \object{B273A} to be $\approx3.5\,{\AA}$, which suggesting a spectral type of G4/G5 for a giant and G2/G3 for a supergiant when following Fig.~2 in \citet{2012A&A...539A.100G}. Alternatively we checked the EW of CO bandhead for the main sequence stars catalogued by \citet{1997ApJS..111..445W} and found that \object{B273A}'s CO bandhead strength is between that of a G8\,V\,(\object{HR\,4496}) and a K2\,V\,(\object{HR\,1084}) star. We could not further constrain the spectral type of \object{B273A} without other characteristic lines at shorter wavelengths.

A fit example is CEN\,34, a mid-G supergiant located along the LOS toward M\,17, is classified to be background with respect to M\,17 with age between 50--100 Myr \citep{2013A&A...557A..51C}. Upon the consideration of \object{B273A}'s apparent brightness at short wavelengths, the object should not be in the foreground of \object{M\,17}. Thus we can rule out the possibility of an early-K dwarf which is only visible at shorter distance. On the other hand, the solution of early-G supergiant puts \object{B273A} in the backside of \object{M\,17}. However, the large amount of molecular gas along the LOS toward the \object{M\,17 UC1\,--\,IRS5} region can almost obscure any background object. Therefore, we keep the solution of mid-G giant as the spectral type of \object{B273A}.

\subsection{Extinction and Luminosity\label{Sec:Ei}}
In order to place \object{IRS5A} and \object{B273A} in the Hertzsprung-Russel diagram~(HRD), one has to determine their luminosity. In principle, the already constrained spectral type of the source would give some hints to derive the extinction toward the source, because the temperature-dependent intrinsic color is reddened by a certain amount of extinction to the observed color. For instance, a general manner of determining the extinction for B stars is applying the short wavelength colors~(e.g. $V-R$ vs $I-R$) rather than the near-IR colors (i.e., $J-H$ vs $H-K$), on consideration that the former ones are more sensitive to the temperatures of early-type stars. In the cases of YSOs, the short wavelength colors are the only reliable way to derive the extinction value because the near-IR colors are usually affected by the strong near-IR excess emission which is emitted by the hot circumstellar material. 

\begin{figure}
\centering
\resizebox{\hsize}{!}{\includegraphics{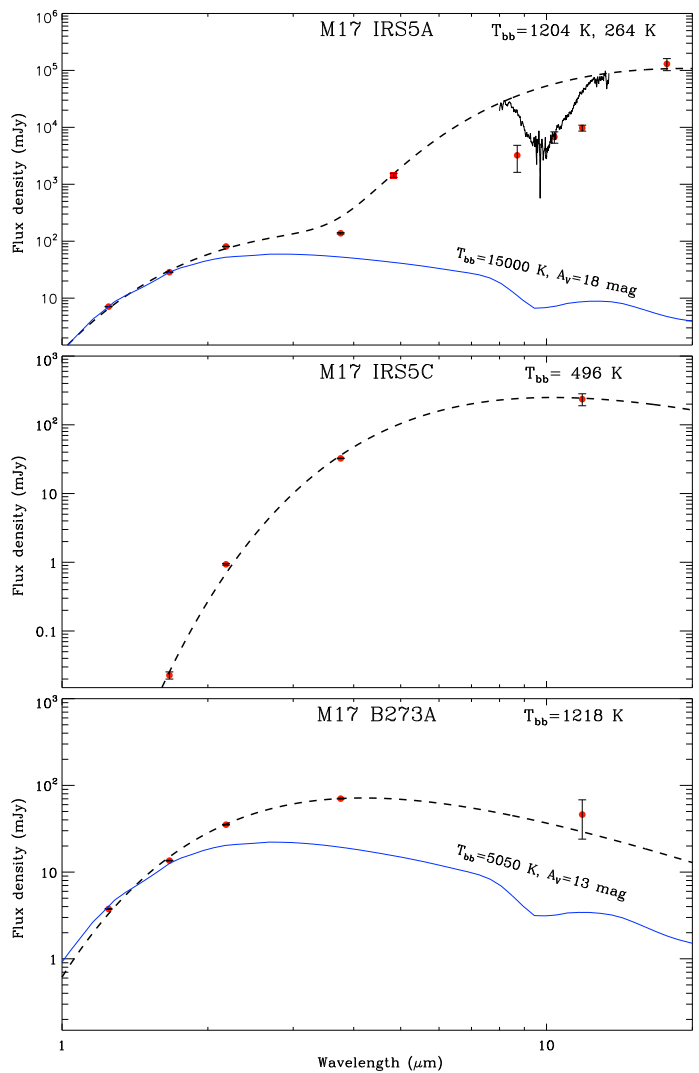}}
\caption{SEDs constructed based on the flux measurements summarized in Table~\ref{Tbl:Ph} and Table~\ref{Tbl:Mir} (filled circles), and the $M$-band point by ISAAC \citep[filled square;][]{2008Hoffmeister}. For \object{IRS5A}, the TIMMI2 spectrum at around $9.7\,\mu$m is denoted by the solid curve in the top panel. The temperature of the blackbody (in blue) corresponds to the proper spectral type according to \citet{2000asqu.book.....C}. }
\label{Fig:Sed}
\end{figure}

The absence of hydrogen emission lines in the spectra of \object{IRS5A} and \object{B273A} suggests that there is no substantial accretion activity. We speculate that the $J-H$ color is not severely affected by dust thermal emission due to the absence of hot accretion tunnel flows. Following the extinction law at near-IR from \citet{1989ApJ...345..245C} and the specific value of $R_V=3.9$ toward M\,17 from \citet{2008ApJ...686..310H}, the $A_V$ values of \object{IRS5A} and \object{B273A} are estimated from their apparent $J-H$ colors on the basis of the intrinsic $J-H$ colors at certain temperatures \citep{2013ApJS..208....9P}. The reliability of the obtained extinction is tested by reddening a blackbody of the proper temperature with the same $A_V$ value. Note that the \citet{1989ApJ...345..245C} near-IR extinction law is invalid in the $5-20\,\mu$m range, hence the extinction law used by \citet{2007ApJS..169..328R} that works in the $1-20\,\mu$m range is employed when reddening the SEDs. In the $1-3\,\mu$m range, the extinction law used by \citet{2007ApJS..169..328R} is equivalent to the \citet{1989ApJ...345..245C} extinction law when $R_V=3.1$.
The reddened blackbody is shown in blue in Fig.~\ref{Fig:Sed}. The blue blackbody curve matches the bluer part of the stellar SED for both objects. On the other hand, \object{IRS5A} displays a deep 9.7\,$\mu$m silicate absorption feature. Using the empirical relation for silicate optical depth $A_V=(18\pm1)\,\tau_{9.7}$ \citep{2003dge..conf.....W}, \object{IRS5A}'s $A_V$ is estimated as $\sim41$ mag from $\tau_{9.7}\sim2.3$. Note that $A_V$ of \object{IRS5A} is even higher than that ($\sim40$\,mag) of \object{UC1} (N+07), while the latter is redder than the former. Such a contradiction even could not be explained by the difference in spectral type between them. Alternatively, if we adopt $A_V/\tau_{9.7}=9-19$ \citep{1985MNRAS.215..425R}, $A_V$ of \object{IRS5A} is hence between $21-44$\,mag. The $A_V$ based on $\tau_{9.7}$ has a large error which originates from the various $A_V/\tau_{9.7}$ ratios. The $A_V$ value of \object{IRS5A} calculated from its $J-H$ color agrees with the upper limit from optical data. \object{IRS5A} and \object{B273A}'s $A_V$ values derived by their $J-H$ colors are adopted in this paper (see Table~\ref{Tbl:Lum}). 

The visual extinction $A_V$ evaluated above is then converted to $A_J$ at $J$-band following the relation $A_J\approx0.27\,A_V$ when $R_V=3.9$. Further, the $J$-band absolute magnitude ($M_J$) can be estimated for both \object{IRS5A} and \object{B273A} through distance modulus. For \object{IRS5A}, the value of $M_J$ is $-3$\,mag. Referring to \citet{2013ApJS..208....9P}, the temperature of \object{IRS5A} ($\sim15000\,\mathrm{K}$) corresponds to a $J$-band bolometric correction factor ($\mathrm{BC}_J$) of $-1.6$\,mag. Hence, the bolometric magnitude, $M_\mathrm{bol}$, of \object{IRS5A} is estimated to be $-4.6$\,mag, corresponding to a bolometric luminosity $\approx5500\,L_{\sun}$. Here the solar $M_\mathrm{bol}$ is adopted as $+4.75$\,mag. Similarly, we have $M_J\approx-1.0$\,mag, and $\mathrm{BC}_J\approx1.25$ for \object{B273A} ($T_\mathrm{eff}\sim 5200\,\mathrm{K}$). $M_\mathrm{bol}\approx0.25$ of \object{B273A} is equal to a bolometric luminosity of $63\,L_{\sun}$.

The flux measurements from near- to mid-IR wavelengths provide an observational approach to evaluate the luminosity. With the new flux measurements in the $1-20\,\mu$m range presented in this paper, we are able to update the IR luminosity of \object{IRS5A}, and derive the IR luminosity for \object{IRS5C} and \object{B273A}. As shown in Fig.~\ref{Fig:Sed}, we employ the simple blackbody curve to fit the infrared flux of the three objects. The significant absorption dip around $10~\mu$m in \object{IRS5A}'s SED corresponds to the $9.7\,\mu$m silicate absorption which is confirmed by the TIMMI2 spectrum. Hence the flux measurements at wavelengths around $9.7\,\mu$m are not used in fitting the SED of \object{IRS5A}. However, for the SEDs of \object{IRS5C} and \object{B273A}, because the $11.85\,\mu$m flux is the only available measurement in mid-IR, we still use this flux measurement in fitting. The $9.7\,\mu$m silicate absorption should be taken into account when dereddening the SED. Be aware of that the $11.85\,\mu$m flux is affected by the $9.7\,\mu$m silicate absorption feature, thus the blackbody fitting will overestimate the color temperatures of \object{IRS5C} and \object{B273A}. 

The SED of \object{IRS5A} can be well fitted by two blackbody components with color temperatures of $1204\,\mathrm{K}$ and $264\,\mathrm{K}$. The cooler component indicates a large amount of dust grains along the LOS of \object{IRS5A}. The color temperatures of \object{IRS5C} and \object{B273A} are 496\,K and 1218\,K, respectively. The observed IR luminosity in the $1-20\,\mu$m range, $L_\mathrm{IR}$, of these objects are summarized in the third column of Table~\ref{Tbl:Lum}.

\object{IRS5A}'s $L_\mathrm{IR}$ is very close to the values in N+01 and K+02, which 
reported values of 3000\,$L_\sun$ and 2600\,$L_\sun$ for \object{IRS5}, respectively. In their work, the aperture size is larger than the one used in this paper, and \object{IRS5C} is not resolved in their larger beam infrared images. A smaller luminosity derived for \object{IRS5A} in this paper is reasonable. On the other hand, the observed $L_\mathrm{IR}$ of \object{IRS5A} and \object{B273A} are much lower than their bolometric luminosity $L_\mathrm{bol}$ derived above. If the observed flux measurements are extrapolated to longer and shorter wavelengths, and are corrected for the extinction, the $L_\mathrm{bol}$ of \object{IRS5A} and \object{B273A} can be constrained observationally.

The extinction-corrected $L_\mathrm{IR}$ is obtained for \object{IRS5A} and \object{B273A} by using the proper extinction law mentioned above. Furthermore, the $L_\mathrm{bol}$ of \object{IRS5A} and \object{B273A} are derived by extrapolating the best-fitting blackbody to longer and shorter wavelengths, e.g. in the range $0.02-200\,\mu$m. The obtained $L_\mathrm{bol}$ is listed in the fourth column of Table~\ref{Tbl:Lum}. 

The $L_\mathrm{bol}$ of \object{IRS5A} and \object{B273A} derived from SED fitting are compared with the values obtained from the temperature-dependent bolometric correction values. For \object{IRS5A}, the former value is significantly larger than the latter one. This difference is partly caused by the extinction used in dereddening the SED of \object{IRS5A}. \object{IRS5A} shows evidence of circumstellar material, the extinction caused by the circumstellar material is included when dereddening the SEDs. However, the circumstellar extinction does not attenuate the $L_\mathrm{bol}$, but just redistributes the SED because the circumstellar material absorbs the blue photons and re-emits at longer wavelengths if assuming an ideal blackbody. In the case of \object{IRS5A}, the circumstellar extinction might be a small portion of the total extinction, upon the consideration that the foreground extinction toward the southwestern bar is suggested to be on the order of 10\,mag or even higher \citep[e.g.][]{2007ApJ...658.1119P,2013ApJ...774L..14S}. We recalculated the $L_\mathrm{bol}$ with a smaller $A_V$ value by 1 mag difference, and got smaller $L_\mathrm{bol}$ of $5.8\times10^3\,L_\sun$ for \object{IRS5A}, which agrees much better with the value of $5.5\times10^3\,L_\sun$. Therefore, we take the $L_\mathrm{bol}$ values in Table~\ref{Tbl:Lum} as the upper limits, and the extinction-corrected $L_\mathrm{IR}$ as the lower limits in the following discussions.

\begin{table}
\caption{Infrared luminosity of the YSOs in the \object{M\,17 UC1\,--\,IRS5} region}
 \centering
  \begin{tabular}{l c c c }    
\hline\hline
      &  $A_V$ &  $L_\mathrm{IR}$ \tablefootmark{a}  & $L_\mathrm{bol}$  \\
      Object     & (mag) & ($L_{\sun}$) & ($L_{\sun}$)   \\
   \hline 
   
    M\,17 B273A  & 13 & 10 (44) & 71 \\
   M\,17 UC1 \tablefootmark{b} & 40 & $2.5\times10^3$ ($1.1\times10^4$) & $3.0\times10^4$\\
    M\,17 IRS5A  & 18 & $2.0\times10^3$ ($3.0\times10^3$) & $8.6\times10^3$\\
   M\,17 IRS5C  & - & 13 & - \\

 \hline
\end{tabular} 
\tablefoot{\\\tablefoottext{a}{Values in brackets are extinction-corrected $L_\mathrm{IR}$.}\\\tablefoottext{b}{Values adopted from N+07.}}
\label{Tbl:Lum}
\end{table}

%


%

\section{Discussion\label{Sec:Di}}
\subsection{Multiplicity \label{Sec:Mul}}
With the high angular resolution infrared images toward M\,17 UC1\,--\,IRS5 region, the companions of the luminous infrared objects (IRS5, B273) are resolved for the first time. According to the $HKL$ color-color diagram, the two components of the \object{B273} system both show $L$-band excess emission. Thus, we suggest that \object{B273A} and \object{B273B} form a young binary system with a projected separation of about 2200\,AU. Three of the five components of \object{IRS5} show $L$-band excess emission. In light of \object{IRS5D}'s position in Fig.~\ref{Fig:Ccd}, we also regard it as a member of \object{IRS5} system taking into account the photometric errors, and suggest \object{IRS5} might be a young quadruple system.

\begin{figure*}
\centering
\includegraphics[width=0.45\textwidth,angle=90]{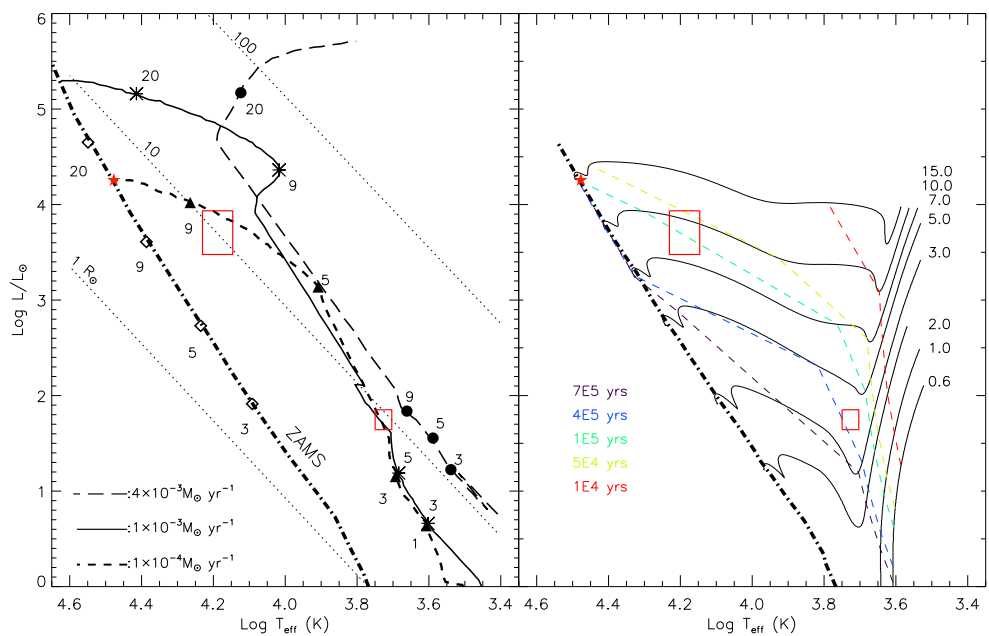}
\caption{Positions of the YSOs in the HRD with well constrained temperatures and luminosities.
Red rectangles outline the parameter space of \object{IRS5A} (the upper one) and \object{B273A} (the lower one), and the red asterisk at upper left marks \object{UC1}.
For \object{IRS5A} and \object{B273A}, the conversion from spectral types to effective temperatures and bolometric luminosities is referred to Sect.~\ref{Sec:Ei}, while \object{UC1} is assumed to be a ZAMS B0.5 star in light of the HCHII region (N+07 and references therein). Left panel: positions of the YSOs compared with the model of massive star formation via disk accretion~(H+10). The symbols along the evolutionary loci at various accretion rates and the ZAMS locus denote the mass at each position. Right panel: positions of the YSOs placed against the PMS evolutionary tracks of different masses~(CA+12). Isochrones of five ages are drawn as the dashed lines in different colors.}
\label{Fig:hrd}
\end{figure*}

\subsection{Structures of the circumstellar material \label{Sec:Cir}}
The large mid-IR excess emission of \object{IRS5A} and \object{UC1} previously reported by N+01 and K+02 indicate circumstellar material for both objects. In addition, from a near-IR polarization view (CZ+12), the high-level polarization individually associated with \object{IRS5A} and \object{UC1} also traces dust grains physically linked with the two objects. Specifically, the almost edge-on circumstellar disk of \object{UC1} found by N+07 produces the high-level $K_s$-band polarization vectors parallel to the disk. The polarization pattern of the IRN illuminated by \object{IRS5A}, however, implies a quite different orientation than \object{UC1}. E.g. the inclination angle of \object{IRS5A}'s IRN estimated in CZ+12 is $\sim50\degr$ from the plane of the sky.

Fig.~\ref{Fig:PolH2} presents perfect geometric match between the H$_2$ emission and polarized light, which explains the physical carriers of the polarized light. The H$_2$ emission residing just beside \object{UC1} is exactly aligned with the elongated feature extending to northwest, which is suggested to coincide with the dust thermal emission originated in the circumstellar disk of \object{UC1} (see Sect.~\ref{Sec:Mir}). We speculate that such H$_2$ emission is related with the circumstellar disk of \object{UC1}, and actually traces the warm H$_2$ gas in \object{UC1}'s disk. The dust grains in \object{UC1}'s disk scatter the incident starlight, and produce the polarized light seen at near-IR. The H$_2$ emission filament related to \object{UC1}'s disk is one of the three H$_2$ emission filaments that are parallel to each other, and parallel to the IF. The H$_2$ filament at the midway between the IF and \object{UC1} is spatially coinciding with the bar-like emission feature seen in the NACO/$K$ image (see Sect.~\ref{Sec:En} and Fig.~\ref{Fig:Pol}). Moreover, the area covered by this bar-like emission in turn is also occupied by the south-north mid-IR extended emission originated from \object{UC1}. This south-north extended emission seen in mid-IR and the coordinated IRN seen in near-IR polarized light together suggest a south-north outflow from \object{UC1}. The bar-like emission seems to arise at the northeastern side of this outflow because this side is facing to the IF. The IF northeast of the outflow suppresses the outflowing material on the northeastern side, and consequently forms a sheet of gas with higher temperature than the rest of the outflow. The temperature difference between the sheet at outflow's northeastern side and the rest of the outflow  explains well the different morphology of the outflow seen between near-IR and mid-IR
wavelengths. Moreover, the coordinated emitting H$_2$ gas at the northeastern side of the outflow is suggested to be excited at a temperature of $\sim600\,\mathrm{K}$ (see Sect.~\ref{Sec:H2}), while the rest of the outflow is suggested to have thermal temperature around $100\,\mathrm{K}$ heated by an early-B outflow driving object \citep{2011ApJ...726...97L}. The north-south outflow of \object{UC1} is most likely the blue-shifted lobe, with the red-shifted lobe invisible because of the severe attenuation.

The coordinated emitting H$_2$ gas surrounding \object{IRS5A} and the IRN illuminated by the same object indicates a distribution of gas and dust embracing \object{IRS5A} in the north with an arc-like shape. The circular shape of the IRN illuminated by \object{IRS5A} reminds us that the southern half would be associated with H$_2$ gas either, although the southern half is not fully covered by the SINFONI data in its small FOV. The temperature of this material distribution is $\sim600\,\mathrm{K}$, similar to the associated emitting H$_2$ gas. With the high-resolution VISIR $11.85\,\mu$m image, we expect a tight relation between the diffuse mid-IR emission and the IRN, just like what has been observed for \object{UC1}. However, the diffuse mid-IR emission is only observed at the midway between \object{IRS5A} and \object{UC1}, where merely a portion of the IRN is spatially associated (see Sect.~\ref{Sec:Mir}). Indeed, this mismatch can be the result of two separate material distributions that are just along the same LOS; the dust temperature responsible for the mid-IR emission of \object{IRS5A} is modeled to be around 264\,K, while the dust temperature linked with the IRN and H$_2$ emission is fitted to be around 575\,K. The two lobes in the core region of \object{IRS5A} stack with the inner circle of the arc-like material along the LOS. The different morphology of \object{IRS5A} seen at near-IR and mid-IR rise the question if the mid-IR emission comes from the area close to \object{IRS5A}, or from its envelope or outflow. The dark lane separating the two mid-IR lobes can be somehow explained by a nearly edge-on cold disk that absorbs the mid-IR emission from backside. The proposed cold disk should have impacts at near-IR bands, such as a silhouette dividing the object into two lobes as well, very similar to \object{UC1}. Nevertheless, \object{IRS5A}'s near-IR images do not show any sign of silhouette disk. The extinction of \object{IRS5A} is another strong issue against the assumption of a nearly edge-on cold disk, because the circumstellar extinction is only a small portion of the total extinction of \object{IRS5A}. We propose that the mid-IR emission actually arises from the outflow whose axis is almost along the LOS of \object{IRS5A}. The two lobes seen in mid-IR correspond to the outflow walls at two sides. In view of the severe attenuation at near-IR bands and at around $10\,\mu$m \citep{2009ApJ...690..496C}, the H$_2$ emission filament is the blue-shifted lobe. Meanwhile the outflowing material traced by the mid-IR emission might be blue-shifted, because of the attenuation issue too.

We note from Fig.~\ref{Fig:H2ext} that \object{IRS5C} is closely associated with H$_2$ 1$-$0\,S(1) emission. Indeed, \object{IRS5C} is embraced by H$_2$ 1$-$0\,S(1) emission which shows an elliptical shape orientated at P.A. $\sim 65\degr$. The bulk of H$_2$ emission is shifted northeast to \object{IRS5C}. As described in Sect.~\ref{Sec:IRE}, \object{IRS5C} is coinciding with 22-GHz H$_2$O maser which is located $0\farcs5$ northwest to \object{IRS5C}. The H$_2$ 1$-$0\,S(1) emission associated with \object{IRS5C} resembles that found for some Herbig Ae/Be stars in their circumstellar disks, where molecular H$_2$ gas is heated by the Ae/Be stars themselves \citep{2011A&A...533A..39C}. The strong excess emission at $L$-band (this work) and mid-IR indicates the presence of circumstellar material surrounding \object{IRS5C}. Hence it is plausible that H$_2$ gas among the warm circumstellar material thermally emits H$_2$ 1$-$0\,S(1) emission detected here. However, unlike the H$_2$ 1$-$0\,S(1) emission detected in the circumstellar disks of \object{HD 97048} and \object{HD 100546} which originates at radii on the order of $\sim 10^1-10^2\,\mathrm{AU}$, the size of the H$_2$ 1$-$0\,S(1) emission associated with \object{IRS5C} is about 1600\,AU in diameter. H$_2$ emission occurring at large distances requests massive stars other than Herbig Ae/Be stars to maintain sufficient radiation field. Nevertheless, the other possibility is that the H$_2$ emission is externally excited by the FUV photons from the ionizing OB stars in M\,17. In light of the excitation mechanism of the H$_2$ emission associated with other objects, the latter explanation is more plausible and reliable. To explain the location of the H$_2$ emission, we propose a nearly edge-on flared disk with radius of 1600\,AU to approximately match the shape of H$_2$ emission. In addition, the 22-GHz H$_2$O maser resides $0\farcs5$ away along the axis of the circumstellar disk of \object{IRS5C}. In some cases, the H$_2$O masers can be excited in the shock fronts where the outflowing gas from high-mass YSO strongly interacts with the surrounding molecular gas \citep[e.g.][]{2014MNRAS.437.3803T}. Considering the position of the H$_2$O maser with respect to the proposed disk, and the disk's size, we interpret that \object{IRS5C} is a deeply embedded high-mass protostar with a nearly edge-on flared disk. \object{IRS5C} does not show radio free-free emission at 1.3\,cm \citep{1998ApJ...500..302J}, which implies that \object{IRS5C} is still at an early stage.

\subsection{Evolutionary Stages of YSOs\label{Sec:Es}}
With the known temperature and luminosity for \object{IRS5A} and \object{B273A}, the locations of these two objects in the HRD will reveal their evolutionary stages when compared with numerical models for star formation. The model of H+10 specifically considers the evolution of a massive protostar via disk accretion, while that of \citet{2012A&A...541A.113C} models the PMS evolution purely tuned by the gravitational contraction without mass feeding, i.e. non-accretion model.

Both models indicate that \object{IRS5A} is still experiencing PMS stage. Using the accretion model H+10 to estimate \object{IRS5A}'s mass and age is problematic, due to the constant accretion rate and fixed initial condition used in H+10. If referring to the non-accretion PMS model, \object{IRS5A}'s mass is currently about $9\,M_{\sun}$ and will not change much when \object{IRS5A} is evolving toward ZAMS. And a non-accretion 0.1\,Myr isochrone passes by \object{IRS5A}'s position in the HRD. Note that non-accretion model simplifies the evolution of protostars, whose accretion histories, initial radii and thermal efficiencies can strongly alter their positions in the HRD. Nevertheless, very young age estimates ($<1\,$Myr) for stars hotter than 3500\,K based on non-accretion model are reliable \citep{2011ApJ...738..140H}. Therefore, we suggest that the age estimate for \object{IRS5A} is reliable.


However, the circumstellar material encircling \object{IRS5A} that is found to be the bipolar molecular outflow driven by itself, and the lack of accretion tracers both doubt that \object{IRS5A} would continue accretion to gain mass. The peculiarity of this object is that it currently does not show any tracer of accretion funnel flow, which imply either periodic accretion or feedback (radiation pressure, outward mass flow) halted accretion. Both mechanisms are thought to have impact on the final mass of high-mass protostar, hence are crucial to understand the formation of high-mass protostar. Otherwise, the detection limit of the SINFONI data could also prevent resolving weak hydrogen emission lines (see the discussion in Sect.~\ref{Sec:Fhc}).  Follow-up investigations for the gas kinematic in the close vicinity of \object{IRS5A} will contribute to determine the gas motion and consequently clarify the evolution trend via constraining the accretion status. Such follow-up studies need high spatial resolution and high spectral resolution observations, which can be implemented by ALMA (Atacama Large Millimeter/submillimeter Array).

\object{B273A} is located in the right-lower area of HRD. Model H+10 interprets mass gaining for \object{B273A}, however, it seems to lack hydrogen recombination lines. The excess emission at $L$-band and mid-IR both indicate the presence of circumstellar material, e.g. a circumstellar disk. Hence, we propose phases of periodic accretion or halted accretion for \object{B273A}, while the emission line strength below the detection limit could be plausible too. In the other case, the location of \object{B273A} in the right panel of Fig.~\ref{Fig:hrd} suggests a mass $\sim4\,M_{\sun}$ and an age $\sim4\times10^5\,\mathrm{yrs}$. As a similar case to \object{IRS5A}, we believe that \object{B273A}'s age estimate based on the non-accretion PMS model is reliable, and \object{B273A} is still at the first-half stage of its entire evolution life toward ZAMS.

\subsection{Feedback on accretion \label{Sec:Fhc}}
One may argue that the spectra of \object{IRS5A} and \object{B273A} have a low resolution ($R\sim1500$) and not high enough S/N, which leads to the 'fake' non-detection in case of weak Br$\gamma$ emission. Based on the spectral resolution ($\mathrm{FWHM}\approx3$ pixels) and continuum's noise level, we estimated that the upper limit of Br$\gamma$ emission flux is $\sim2.2\times10^{-15}\,\mathrm{ergs}\,\mathrm{cm}^{-2}\,\mathrm{s}^{-1}$ for \object{IRS5A}, and $7.9\times10^{-16}\,\mathrm{ergs}\,\mathrm{cm}^{-2}\,\mathrm{s}^{-1}$ for \object{B273A}; with the extinction values of this two objects shown in Table~\ref{Tbl:Lum}, the extinction-corrected Br$\gamma$ line flux, $f_{\mathrm{Br}\gamma}$, is $\sim1.3\times10^{-14}\,\mathrm{ergs}\,\mathrm{cm}^{-2}\,\mathrm{s}^{-1}$ for \object{IRS5A}, and $\sim2.8\times10^{-15}\,\mathrm{ergs}\,\mathrm{cm}^{-2}\,\mathrm{s}^{-1}$ for \object{B273A}. Further the luminosity of Br$\gamma$ emission line was computed as $L_{\mathrm{Br}\gamma}=4\,\pi\,d^2\,f_{\mathrm{Br}\gamma}$, where $d$ is the distance of M\,17. The Br$\gamma$ line luminosity is found to be tightly correlated with the accretion luminosity, $L_\mathrm{acc}$, for a statistical sample of low-mass YSOs \citep[e.g. ][]{2011A&A...535A..99M,2012A&A...548A..56R,2014A&A...561A...2A}. For high-mass YSOs, the correlation between $L_{\mathrm{Br}\gamma}$ and $L_\mathrm{acc}$ is not well established due to the lack of a statistics meaningful sample. As an approximation, we used the $\log L_{\mathrm{Br}\gamma}-\log L_\mathrm{acc}$ relation reported in \citet{2014A&A...561A...2A} to estimate the $L_\mathrm{acc}$, which is $\sim2.2\,L_\sun$ for \object{IRS5A}, and $\sim0.38\,L_\sun$ for \object{B273A}. From the positions of this two objects in the HRD, we estimated a radius of $\sim10\,R_\sun$ for both of them. Together with the masses of the two objects mentioned above, the formula $$\dot{M}_\mathrm{acc}\approx1.25\,\frac{L_\mathrm{acc}\,R_\ast}{G\,M_\ast}$$ gives the upper limit of $\dot{M}_\mathrm{acc}$, which is $\sim1.1\times10^{-7}\,M_{\sun}\,\mathrm{yr}^{-1}$ for \object{IRS5A}, and $\sim3.8\times10^{-8}\,M_{\sun}\,\mathrm{yr}^{-1}$ for \object{B273A}.

High-mass stars can arrive at ZAMS stage still during their accretion phase. However, the enormous radiation pressure and ionizing photons will act on the accretion flow and dissipate the circumstellar material. Thus, the timescale of the accretion phase for a high-mass protostar will strongly depend on the stellar mass, which in turn is constrained by accretion rate. Such timescale is far to be known for high-mass protostars. \object{UC1}'s disk suggests that this B0.5 ZAMS star might be still in its accretion phase. A plausible evidence supporting accretion comes from the Br$\gamma$ emission at the position of \object{UC1}'s southwest lobe (Fig.~\ref{Fig:Lmp}). Nevertheless, Br$\gamma$ emission can be produced in the HCHII region associated with \object{UC1}. For an HCHII region, the dynamics and morphology of the ionized gas are determined by the ratio of the ionization radius, $R_i$, and the gravitational bound radius, $R_b$, defined by $G\,M_\ast/2\,c_s^2$, where $M_\ast$ is the stellar mass and $c_s$ is the sound speed \citep{2007ApJ...666..976K}. With $M_\ast=15\,M_\sun$ and $c_s=10\,\mathrm{km\,s}^{-1}$ (at the temperature of $10^4$\,K for ionized gas), $R_b$ is 66\,AU. Since the Br$\gamma$ emission must come from the ionized gas of the HCHII region, the Br$\gamma$ line map in Fig.~\ref{Fig:Lmp} shows the morphology of the HCHII region around UC1. The distance from the obscured centroid of \object{UC1} to the southwestern Br$\gamma$ emission knot is taken as the value of $R_i$, which is $\sim0\farcs25$, or $\sim500$\,AU. Thus, for \object{UC1}, $R_i \gg R_b$, corresponding to a stage of HCHII region that the ionized gas moves outward to form an outflow \citep{2007ApJ...666..976K}. In this picture, the accretion is confined to a narrow range of angles close to the mid-plane of the disk and is close to termination. Without the observational evidence of the gas dynamics in the inner most region of \object{UC1}, the accretion status of \object{UC1} is controversial. 

Due to the large inclination angle of \object{IRS5A} from the plane of the sky, near-IR emission lines produced in the accretion flow (if exists) can be seen without the severe attenuation by the disk. In fact, \object{IRS5A} does not show any emission line indicating accretion. We have proposed two reasons for the absence of accretion indicators in Sect.~\ref{Sec:Es}. If in the case of non-accretion phase, the evolutionary stage of \object{IRS5A} would imply that accretion could even be halted before high-mass protostar arriving at ZAMS. In view of the lower temperature at the early stage of high-mass protostar, radiation feedback is not as strong as for a ZAMS star, thus outflow activities are additionally required to be responsible for the reversed accretion. In a large picture, the expanding \ion{H}{ii} region of M\,17 might also contribute to the dissipation of accretion tunnel flows. The deduced timescale of accretion phase is thus $\sim1\times10^5\,\mathrm{yrs}$, on the same order of high-mass protostar's age. On the other hand, the probable periodic accretion process of \object{IRS5A}  would suggest a timescale of accretion phase much longer. The determination of the gas dynamics in the inner most area of \object{IRS5A} will disentangle this confusing puzzle.

Interestingly, the absence of hydrogen recombination lines in \object{B273A}'s spectrum implies a terminated-accretion phase. An intermediate-mass YSO with an age $\sim4\times10^5\,\mathrm{yrs}$ like \object{B273A} is most likely to show accretion activity, because YSOs at later evolutionary stages such as intermediate-mass T Tauri stars ($1.5-4\,M_\sun$, $1-10\,\mathrm{Myr}$) were observed to show Br$\gamma$ emission even in low-resolution ($R\sim800$) near-IR spectra \citep{2004AJ....128.1294C}. This discrepancy between the observation and the predication for \object{B273A} indicates additional feedback to dissipate the accretion funnel flows. The exact spatial coincident between \object{B273A} and the IF implies that the expanding \ion{H}{ii} region driven by the ionizing photons is the plausible mechanism of external feedback.

From the current masses and ages of these two objects, \object{IRS5A} and \object{B273A} require mass accretion rate of $9\times10^{-5}\,M_{\sun}\,\mathrm{yr}^{-1}$ and $1\times10^{-5}\,M_{\sun}\,\mathrm{yr}^{-1}$, respectively. Even in the possibility of weak Br$\gamma$, we still note a dramatically drop of $\dot{M}$ for both objects. This could be also the result of feedback from stellar activities and expanding \ion{H}{ii} region. For the two low-mass YSOs, \object{B273B} and \object{IRS5B}, we could similarly estimate that the upper limit of their mass accretion rates is definitely lower than that of \object{B273A} , i.e.~$\sim3.8\times10^{-8}\,M_{\sun}\,\mathrm{yr}^{-1}$. The typical mass accretion rate of low-mass YSOs is on the order of $1\times10^{-8}\,M_{\sun}\,\mathrm{yr}^{-1}$ \citep[e.g.][]{2013A&A...549A..15F,2014A&A...561A...2A}, which is comparable to the detection limit of \object{B273B} and \object{IRS5B}'s spectra. The lack of Br$\gamma$ emission line in the spectra of low-mass YSOs (e.g., \object{B273B}, \object{IRS5B}) is reminiscent of non-accretion phase, however, we cannot rule out the possibility that their accretion indicators could be revealed by higher resolution spectroscopic observation. In the case of non-accretion phase for these low-mass YSOs, it seems to remind us that such external radiation feedback might be important in the disk clearing process for the YSOs in massive star forming regions.


\subsection{Properties of M\,17 SW \label{Sec:PDR}}
Previous studies of M\,17 SW speculated that a dense PDR with high-density clumps and low-density gas to match the observed properties of M\,17 SW \citep[][]{1992ApJ...390..499M,2010A&A...510A..87P,2012A&A...542L..13P}. However, these studies are based on far-IR and radio line observations, and thus cannot resolve structures smaller than several arcseconds. The near-IR H$_2$ emission detected in the \object{M\,17 UC1\,--\,IRS5} region with extremely high spatial resolution (100 mas) set a entirely new picture for the warmest part of M\,17 SW. The column densities of several H$_2$ $v=1-0$ ro-vibrational lines show that H$_2$ gas mostly concentrates in filaments in the range $0.02-0.05\,\mathrm{pc}$. The H$_2$ line ratios, as been discussed in Sect.~\ref{Sec:H2}, refer to a gas density of $>10^5\,\mathrm{cm}^{-3}$ if they are compared to PDR models \citep{1990ApJ...352..625B}. The high-density nature of \object{M\,17 SW} is also evidenced by the molecular gas persisting just west of the IF \citep{2012A&A...542L..13P}. This can be interpreted as the result of self-shielding of molecules in high-density PDR which can move the transition regions for $\ion{H}{i}/\mathrm{H}_2$ close to the surface of the cloud \citep{1990ApJ...352..625B}. Complementally, the H$_2$ gas temperature characterized by the ro-vibrational diagram is in the range $500-750\,\mathrm{K}$. In contrast, \ion{H}{i} observations toward M\,17 specifically showed that \ion{H}{i} number density toward the \object{M\,17 UC1\,--\,IRS5} region is in the range $3.7\times10^3-1.5\times10^4\,\mathrm{cm}^{-3}$, which corresponds to  \ion{H}{i} spin temperature between 50\,K and 200\,K in many directions toward \object{M\,17} \citep{1999ApJ...515..304B}. Therefore, the H$_2$ filaments presented here confirm for the first time that dense clumps with number density two orders of magnitude higher and temperatures higher than ambient atomic gas can exist down to 10$^{-2}$\,pc scale inside a dense PDR like M\,17 SW, which was previously suggested being clumpy \citep{1992ApJ...390..499M}.

\section{Summary and conclusions\label{Sec:Co}}
In this paper we presented diffraction-limited near- to mid-IR images and SINFONI integral field spectroscopy at $H+K$ toward the M\,17 UC1\,--\,IRS5 region, which shows emission lines of \ion{He}{i}, \ion{H}{i}, and H$_2$. Our diffraction-limited data reveal new fine structures of this region.  
  \begin{enumerate}
      \item This work complements the work by CZ+12 from a view of much higher angular resolution. The IRN identified in CZ+12 illuminated by \object{IRS5A} and by \object{UC1} are confirmed to trace the molecular outflow driven by the two objects, respectively. Combining the SINFONI H$_2$ line emission map and the mid-IR VISIR image, a blue-shifted outflow lobe is proposed for \object{IRS5A}, with two types of tracers in the form of H$_2$ emission filament and mid-IR emission. The molecular outflow \object{UC1} is running south-north with its blue-shifted lobe merely visible in forms of near-IR polarized light and mid-IR emission. The northeastern side of \object{UC1}'s outflow is suppressed by the shock fronts driven by the expanding \ion{H}{ii} region. Thus a sheet of warm, dense gas forms at the northeastern edge of \object{UC1}'s molecular outflow, and appears in forms of the bar-like emission seen in $K$-band and the H$_2$ emission filament which are both parallel to the ionization front. \\

      \item The uniform line ratios of all H$_2$ emitting areas indicate the same excitation mechanism; the H$_2$ molecules are initially pumped by FUV photons, and are repopulated by the collisional deexcitation in a dense PDR. The H$_2$ gas excitation temperature is estimated to be around 575\,K based on the ro-vibrational diagram of the detected H$_2$ $v=1$ lines. The H$_2$ line ratios are used to probe the properties of the PDR. Comparisons to PDR models suggest clumpy PDR structure with at least two components: warm, dense molecular clumps with $n_\mathrm{H}>10^5\,\mathrm{cm}^{-3}$ and $T\sim575$\,K and atomic gas with $n_\mathrm{H}\sim3.7\times10^3-1.5\times10^4\,\mathrm{cm}^{-3}$ and $T\sim50-200$\,K.\\

      \item  \object{IRS5} might be a young quadruple system containing IRS5A, IRS5B, IRS5C, and IRS5D. Its primary star \object{IRS5A} is classified as a B3--B7\,V/III star with a bolometric luminosity of $3.0-8.6\times10^3\,L_\sun$, confirmed to be a high-mass protostar with mass $\sim 9\,M_{\sun}$ and age $\sim 1\times10^5\,\mathrm{yrs}$, while the three lower mass companions are much less constrained. Particularly, \object{IRS5C} might be a deeply embedded high-mass protostellar object with a dusty disk. The spectral type of \object{B273A} is assigned to G4/5\,III, suggesting an intermediate-mass YSO of $\sim4\,M_{\sun}$ and an age $\sim4\times10^5\,\mathrm{yrs}$ when compared with the non-accretion PMS model. \object{B273A} may have a lower mass companion \object{B273B} of possibly similar age.\\

     \item The Br$\gamma$ emission of \object{UC1} is ambiguous for tracing the accretion because the emission can arise from the HCHII region as well; \object{UC1} might have terminated accretion from a concern on the ratio of $R_i\gg R_b$. The absence of Br$\gamma$ emission in the SINFONI spectra of the other YSOs of various masses implies terminated accretion for them. Even in the case of weak Br$\gamma$ emission that cannot be resolved by the SINFONI data, the steep drop in the accretion rates of \object{IRS5A} and \object{B273A} along the protostellar evolution suggests processes of dissipating accretion funnel flows. The object \object{IRS5A} is unique because it is during a stage when its radiation feedback is not high enough to severely destroy accretion funnel flows or even to terminate the accretion; therefore, the expanding \ion{H}{ii} region and its bipolar outflow both are suggested to contribute simultaneously. For the other YSOs, the expanding \ion{H}{ii} region could be the major mechanism on dissipating/terminating accretion.

   \end{enumerate}

\begin{acknowledgements}
This work is supported by the Strategic Priority Research Program ``The Emergence of Cosmological Structure'' of the Chinese Academy of Sciences, grant No. XDB09000000. Z.J. acknowledges the support by NSFC 11233007. M.F. acknowledges the support by the NSFC through grants 11203081. Z.C. acknowledges Dr. S. Zhang for his IDL routines to make good-look RGB figures. This publication used data products from the Two Micron All Sky Survey, which is a joint project of the University of Massachusetts and the Infrared Processing and Analysis Center/California Institute of Technology, funded by the National Aeronautics and Space Administration and the National Science Foundation. This research made use of the SIMBAD database, operated at the CDS, Strasbourg, France. This research made use of NASA's Astrophysics Data System Bibliographic Services.
\end{acknowledgements}

\bibliographystyle{aa} 
\bibliography{myrefs}

\end{document}